\documentclass[prd,aps,preprint,amsmath,amssymb,eshowkeys,nofootinbib]{revtex4-2}
\usepackage{verbatim,graphics,graphicx,color,slashed,textcomp,bbm,mathdots,multirow,array}
\usepackage[normalem]{ulem} 
\usepackage[colorlinks=true,linkcolor=red,urlcolor=blue,citecolor=blue]{hyperref}

\usepackage{autobreak}
\usepackage{float}

\graphicspath{{figures/}}

\makeatletter

\newcommand{\Rmnum}[1]{\expandafter\@slowromancap\romannumeral #1@}
\makeatother

\usepackage[colorlinks=true,linkcolor=red,urlcolor=blue,citecolor=blue]{hyperref}

\graphicspath{{figures/}}

\begin{document}


\title{Sensitivity of Space-based Gravitational-Wave Interferometers to Ultralight Bosonic Fields and Dark Matter}
\author{Jiang-Chuan Yu$^{a,c}$}
\altaffiliation{Two authors contribute equally.}
\author{Yue-Hui Yao$^{a,c}$}
\altaffiliation{Two authors contribute equally.}
\author{Yong Tang$^{a,b,c,d}$}
\author{Yue-Liang Wu$^{a,b,c,e}$}
\affiliation{\begin{footnotesize}
		${}^a$University of Chinese Academy of Sciences (UCAS), Beijing 100049, China\\
		${}^b$School of Fundamental Physics and Mathematical Sciences, \\
		Hangzhou Institute for Advanced Study, UCAS, Hangzhou 310024, China \\
		${}^c$International Center for Theoretical Physics Asia-Pacific, Beijing/Hangzhou, China \\
		${}^d$National Astronomical Observatories, Chinese Academy of Sciences, Beijing 100101, China\\
		${}^e$Institute of Theoretical Physics, Chinese Academy of Sciences, Beijing 100190, China
		\end{footnotesize}}

\date{\today}

\begin{abstract}
Ultralight bosonic fields~(ULBFs) are predicted by various theories beyond the standard model of particle physics and are viable candidates of cold dark matter. There have been increasing interests to search for the ULBFs in physical and astronomical experiments. In this paper, we investigate the sensitivity of several planned space-based gravitational-wave interferometers to ultralight scalar and vector fields. Using time-delay interferometry~(TDI) to suppress the overwhelming laser frequency noise, we derive the averaged transfer functions of different TDI combinations to scalar and vector fields, and estimate the impacts of bosonic field's velocities. We obtain the sensitivity curves for LISA, Taiji and TianQin, and explore their projected constraints on the couplings between ULBFs and standard model particles, illustrating with the ULBFs as dark matter.

\end{abstract}

\keywords{Dark Matter, Ultralight fields, Gravitational wave, Time-Delay Interferometry}
\maketitle
\newpage

\section{INTRODUCTION}\label{sec:intro}
On September 14, 2015 the LIGO detectors detected the first gravitational-wave signal generated from a binary black hole merger~\cite{LIGOScientific:2016aoc}, which opens the era of gravitational astronomy~\cite{Sathyaprakash:2009xs}. Since then, over ninety gravitational wave events have been detected, which not only brings us many valuable information about the Universe but also shows the incredible ability of interferometry detector. Precision interferometers can measure the relative displacements of two mirrors separated several kilometers apart at subatomic level. Besides gravitational waves, other physical processes may also generate the relative displacement of mirrors (test masses). Thus interferometer gravitational-wave detectors can also be used to search for the signals of those physical processes. 

This idea has been recently explored, for example in~\cite{PhysRevLett.121.061102, PhysRevD.100.123512, Michimura:2020vxn, Morisaki:2020gui}, which propose to use interferometers to search for ultralight dark matter (ULDM)~\cite{Kawasaki:2013ae, Marsh:2015xka, Hui:2016ltb, Hui:2021tkt, Ferreira:2020fam}. Since ULDM particles are very light, typically below $1$~eV, there are numerous ULDM particles in a de Broglie volume (for $m=10^{-17}$~eV, the average number of particles $\sim 10^{51}$). The occupancy number is so large that the quantum fluctuations are negligible and the set of ULDM particles can be well described by classical bosonic fields that oscillate at the Compton frequency. When ULDM is coupled to the SM particles, its oscillation will effectively exert a periodical force on the test masses and induce spatial displacements of them, which may end up with detectable signal in the interferometer. Searching for the ULDM signals in the data of LIGO~\cite{LIGOScientific:2021ffg, Guo:2019ker, Fukusumi:2023kqd} and LISA Pathfinder~\cite{PhysRevD.107.063015} has been conducted recently. Although reporting null results, these works show that high precision interferometers can put stringent constraints on the coupling constants close to or even comparable with constraints given by equivalence principle tests and fifth-force experiments~\cite{Adelberger:2003zx, Schlamminger:2007ht, MICROSCOPE:2022doy, Graham:2015ifn, Grote:2019uvn}. 

In the above works, it was assumed that the ultralight bosonic fields serve as dark matter, therefore the amplitudes of fields are determined by the local dark matter density and the frequency dispersion of fields follows from the velocity dispersion of DM. However, besides DM, ULBFs are widely predicted in physical and cosmological theories for various motivations, and have been actively investigated on different forms~\cite{Peccei.Quinn, Wilczek.1978, Weinberg.1978, Preskill:1982cy, Abbott:1982af, Dine:1982ah, Nelson:2011sf, Graham:2015rva, Ema:2019yrd, Ahmed:2020fhc, Agrawal:2018vin, Tang:2020ovf, Kitajima:2023fun, Kolb:2020fwh, An:2020jmf, Moroi:2020has, Liang:2022gdk, Sun:2021yra, An:2022hhb, Luu:2023rgg, Liu:2021zlt, An:2022hhb, Chen:2019fsq, Tang:2023oid, Grote:2019uvn, Xia:2020apm, Fayet:1980ad, Fayet:1989mq, Fayet:1990wx, Fayet:2017pdp, Wu:2023dnp} since many possible field configurations could exist. In this work, we take a systematic approach and drop the assumption that ULBF makes up the local DM,  which implies the amplitude of field and the velocity of filed are generic. We investigate the ability of interferometers to constrain the combinations of coupling constants and field amplitudes in the more general setup and illustrate with the DM case. In the prospects of the increasing sensitivity of operating interferometer detectors~\cite{LIGOScientific:2014pky, VIRGO:2014yos} and the next generation ground-based detectors like Cosmic Explorer~\cite{Cosmic:2021gyd}, Einstein Telescope~\cite{Punturo:2010zz} and space-based detectors like LISA~\cite{LISA:2017pwj}, Taiji~\cite{Hu:2017mde} and TianQin~\cite{TianQin:2015yph}, we show it is possible to detect weaker ULBF signals or put more stringent constraints on the coupling constants and amplitudes of the fields in the null case by space-based detectors.

In this work, we focus on two kinds of ULBF, scalar and vector, in the mass range $10^{-18} \sim 10^{-13}$~eV, which corresponds to the sensitive band of space-based gravitational-wave interferometers. We investigate their potential to detect ULBF signals by calculating response functions and sensitivity curves of several main time-delay interferometry combinations~\cite{Estabrook:2000ef, Tinto:2004wu}, such as Michelson $X,Y,Z$, Sagnac $\alpha,\beta,\gamma$, fully symmetric Sagnac $\zeta$ and optimal channel $\eta$. Our results show future space-based detectors can provide a better probe in some parameter regions.

This paper is organized as follows. In Sec.~\ref{sec:model}, we present the scalar and vector models and explain how these ultralight bosonic fields affect the motion of test masses. One of the key result is the single-link transfer function of interferometer to the bosonic fields. In Sec.~\ref{sec:transfer}, we construct the sky and polarization averaged transfer functions of different TDI combinations from the single-link transfer function and discuss their asymptotic behaviors in the long wavelength limit and contrast them with the case for gravitational wave. In particular, we show the effects of field velocity on the transfer function. Then we calculate the sensitivity curves of Taiji, LISA and TianQin to ULBF in Sec.~\ref{sec:sensitivity} and illustrate with dark matter case in Sec.~\ref{sec:constraint}. Finally, we give our conclusion in Sec.~\ref{sec:summary}. 

In this paper, we use natural units~($c=\hbar=1$) unless explicitly stated otherwise.

\section{Formalism and Conventions}\label{sec:model}

In this section, we introduce our conventions for scalar and vector fields, and their ways of interacting with ordinary matter. We obtain the equation of motion for the test masses through the action of the system and give the spatial displacements of the test masses, which is essential for subsequent discussions. The physical picture is that the spatial shifts affect the optical lengths of laser beams, which induces the phase difference or equivalently the frequency shift that will be detected by the phasemeters. 

\subsection{Scalar field}
Ultralight scalar fields for spin-0 particles appear in many popular physical and cosmological theories, such as axions, dark matter, dilaton and dark energy models. Here, we do not specify the exact model but only present a general formalism. We denote $\phi$ as a scalar field which linearly couples to SM particle mass terms. Then the effective Lagrangian of the system at the low-energy limit is given by
\begin{equation}
    \mathcal{L} = \mathcal{L}_{\phi} + \mathcal{L}_{SM} + \mathcal{L}_{\phi-SM},
\end{equation}
where $\mathcal{L}_{\phi}$ is the Lagrangian of the scalar field and usually can be expressed by $\mathcal{L}_{\phi}=-\frac{1}{2}\partial_{\mu}\phi \partial^{\mu}\phi - \frac{1}{2}m_{\phi}^{2}\phi^{2}$, $\mathcal{L}_{SM}$ is the Lagrangian of the Standard Model and $\mathcal{L}_{\phi-SM}$ for the interaction between the scalar field and the ordinary matter. The details of the interaction will be presented in Sec.~\ref{sec:constraint}. At the moment, we can work in the low-energy limit and treat the test mass $m$ as a function of $\phi$ and the action of the test mass is given by~\cite{PhysRevD.100.123512}
\begin{equation}
    S = -\int m(\phi) \sqrt{-\eta_{\mu \nu} dx^{\mu} dx^{\nu}}.
\end{equation}
The effective scalar-matter coupling $\alpha$ is encoded in the mass and defined by
\begin{equation}\label{eq:alpha}
    \alpha = \frac{1}{\kappa^{2}m(\phi)}\frac{\partial[\kappa m(\phi)]}{\partial \phi},
\end{equation}
where $\kappa \equiv \sqrt{4\pi G}=\sqrt{4\pi }/M_P$ is the inverse of the Planck mass and used here for a reference scale. 

Without loss of generality, we use a monochromatic plane wave to describe the scalar field as a transient source:
\begin{equation}
    \phi \left( t, \vec{x} \right)= \phi_{\vec{k}} e^{i(\omega t - \vec{k} \cdot \vec{x} + \theta_{0})},
\end{equation}
where $\phi_{\vec{k}}$ is the amplitude of the wave, $\omega$ the de Broglie angular frequency, $\vec{k}$ the momentum vector of the scalar field, and $\theta_{0}$ an initial phase whose exact value does not affect our discussion and can be set to zero. When the scalar field is referred as DM in our galaxy, it has a velocity dispersion, $v \sim 10^{-3}$, which implies that the wave is non-relativistic with $\vec{k} \approx m_{\phi}\vec{v}$ and $\omega \approx m_{\phi}$. We can also extend the case to the relativistic condition, the details will be discussed later. 

Using the Euler-Lagrangian equation and neglecting the subdominant term related to the velocity of test mass, we get the additional acceleration and displacement induced by the oscillation of scalar field:
\begin{equation} 
    a^{i}(t,\vec{x}) =  i \phi_{\vec{k}}\;\alpha \kappa \; k^i e^{i(\omega t - \vec{k} \cdot \vec{x}) }\simeq i\phi_{\vec{k}}\;\alpha \kappa \; k^i e^{im_{\phi}( t - v\hat{k} \cdot \vec{x})}.
\end{equation}
We can integrate the above equations twice and obtain the spatial displacement,
\begin{equation}\label{disvar}
    \delta x^{i}(t,\vec{x}) = -i \alpha \kappa \; \phi_{\vec{k}} \frac{k^{i}}{m_{\phi}^{2}} e^{im_{\phi}( t - v\hat{k} \cdot \vec{x})} 
    =-i\mathcal{M}_{s}\hat{k}^{i} e^{im_{\phi}( t - v\hat{k} \cdot \vec{x})},
\end{equation}
where $\hat{k}$ is the unit vector of $\vec{k}$ and $\mathcal{M}_{s}=\alpha \kappa \phi_{\vec{k}} |\vec{k}|/m_{\phi}^{2}$. 


\subsection{Vector field}
Ultralight vector fields for spin-1 particles are also well motivated theoretically, such as very light scaling gauge field for DM candidate~\cite{Wu:2015wwa, Wu:2017urh, Wu:2022mzr, Tang:2020ovf, Wang:2022ojc}. Vector field can directly couple with the baryon number~(B) or baryon minus lepton number~(B-L) and it will exert an additional force on the test masses, which leads to the periodic displacements of test masses.

The Lagrangian of the massive vector field coupled with B or B-L current $J_{D}$ is given by
\begin{equation}
    \mathcal{L} = -\frac{1}{4}F^{\mu \nu}F_{\mu \nu} +\frac{1}{2}m_{A}^{2}A^{\nu}A_{\nu}-\epsilon_{D} e J^{\nu}_{D}A_{\nu},
\end{equation}
where $F_{\mu \nu} = \partial_{\mu}A_{\nu}-\partial_{\nu}A_{\mu}$, $A^{\mu}$ is the vector field and $m_{A}$ is the mass of vector field. Here we normalize the coupling strength of the vector field $e_{D}$ in terms of the electromagnetic coupling constant $e$ and denote $\epsilon_{D}$ as the ratio $\epsilon_{D} = e_{D}/e$~\cite{PhysRevD.102.102001}.

The spatial components of $A^{\mu}$ are described by 
\begin{equation}
    \vec{A}(t,\vec{x}) = |\vec{A}|\hat{e}_{A}e^{i(\omega t-\vec{k}\cdot \vec{x})} ,
\end{equation}
where $\hat{e}_{A}$ is the unit polarization vector which is parallel to $\vec{A}$. Here we ignore $A^{0}(t,\vec{x})$ since it is much smaller than $A^{i}(t,\vec{x})$ under Lorentz gauge in the non-relativistic limit.
A object carrying B or B-L charge can be accelerated by the electric-like field generated by the vector field. The acceleration acting on the test mass can be written as \cite{PhysRevD.102.102001, PhysRevLett.121.061102}
\begin{align}\label{ac2}
    {a^{i}}(t,\vec{x}) = \epsilon_{D} e \frac{q_{D,j}}{M_{j}} \partial_{t}A^{i}(t,\vec{x})
    \simeq i\epsilon_{D} e \frac{q_{D,j}}{M_{j}} m_{A} |\vec{A}| \hat{e}_{A}^{i} e^{im_{A}(t-v\hat{k}\cdot \vec{x})}.
\end{align}
where $M_{j}$ is the mass of the jth test mass. $q_{D,j}$ is the total baryon number if the vector field is a $U(1)_{B}$ gauge field, and only counts the neutrons in the test masses if it is associated with $U(1)_{B-L}$ gauge symmetry.
We can integrate Eq.~(\ref{ac2}) twice and obtain the spatial displacement of the test mass:
\begin{equation}\label{ac22}
    \delta x^{i}(t,\vec{x}) = -i\epsilon_{D} e \frac{q_{D,j}}{M_{j}} \frac{|\vec{A}|}{m_{A}} \hat{e}_{A}^{i} e^{im_{A}(t-v\hat{k}\cdot \vec{x})}
    = -i \mathcal{M}_{v} \hat{e}_{A}^{i} e^{im_{A}(t-v\hat{k}\cdot \vec{x})},
\end{equation}
where $\mathcal{M}_{v}= \epsilon_{D} e q_{D,j} |\vec{A}|/m_{A}M_{j}$. Comparing it with the scalar case, Eq.~(\ref{disvar}), we can find that the difference is that $\hat{k}^{i} $ in Eq.~(\ref{disvar}) is replaced by $\hat{e}_{A}^{i}$ in Eq.~(\ref{ac22}). Therefore, both transverse and longitudinal modes contribute. 

\section{Transfer function}\label{sec:transfer}
In this section, we discuss how the displacement signals induced by ULBFs manifest in the physical measurements. We establish the necessary formalism to describe the transfer function, including the detector response of a single link and time-delay interferometry. 

\subsection{The detector frame}
\begin{figure*} 
\includegraphics[scale=0.35]{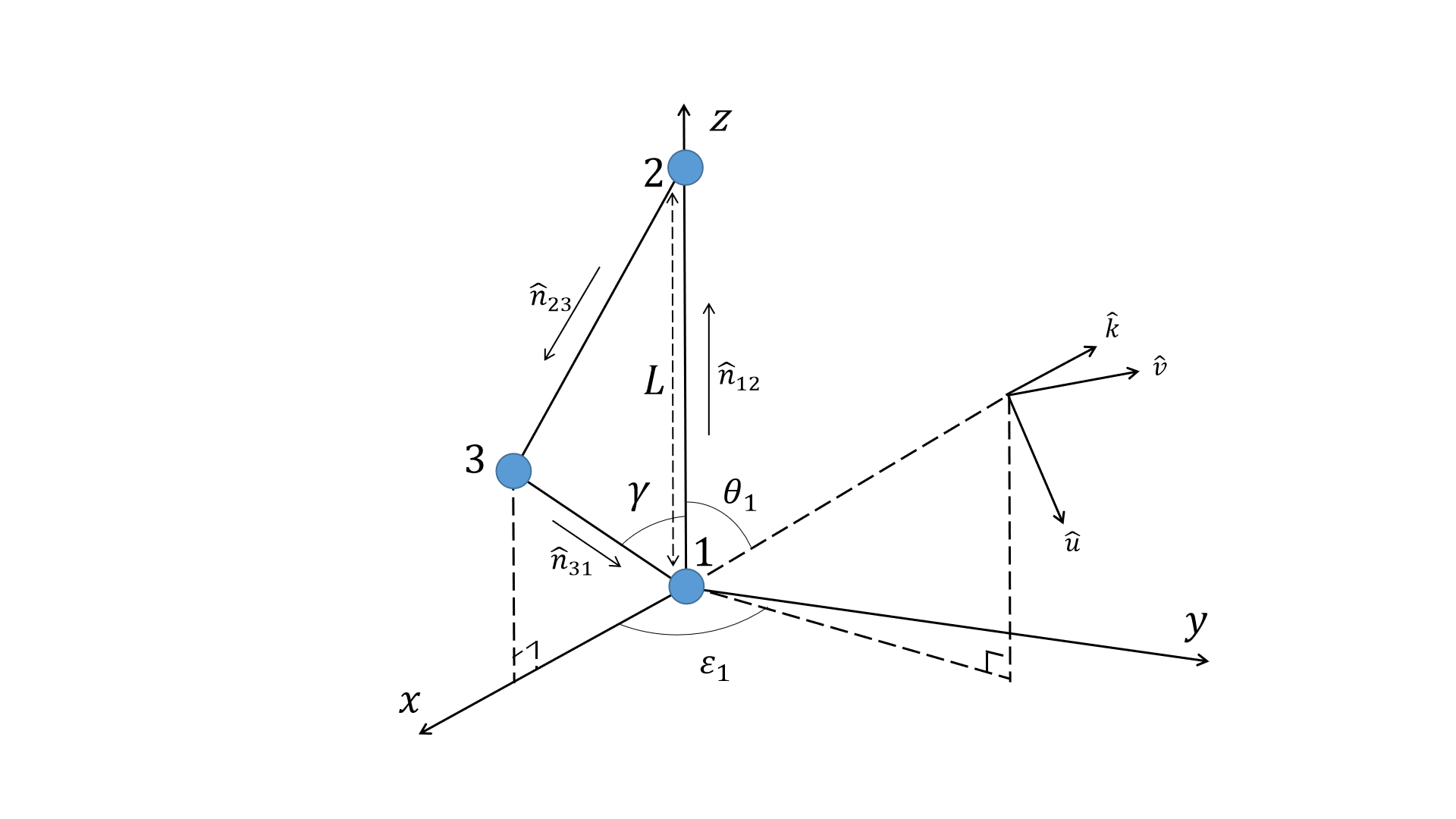}
\caption{Schematic of the detector. 1, 2, 3 represent the labels of spacecrafts. Three spacecrafts are located at $\vec{x}_{1}=(0,0,0)$, $\vec{x}_{2}=(0,0,L)$ , and $ \vec{x}_{3}=(L\sin\gamma,0,L\cos\gamma)$ respectively. The static distance between two spacecrafts is $L$, and the angle between two arms is $\gamma = \frac{\pi}{3}$. $\hat{n}_{rs}$ refers to the unit vector in the direction from $\vec{x}_{r}$ to $\vec{x}_{s}$. The spatial direction of the wave vector $\vec{k}$ is $\hat{k}=(\sin\theta_{1}\cos\epsilon_{1}, \sin\theta_{1}\sin\epsilon_{1},\cos\theta_{1})$, and $\hat{u}=(\cos\theta_{1}\cos\epsilon_{1}, \cos\theta_{1}\sin\epsilon_{1},-\sin\theta_{1})$, $\hat{v}=(-\sin\epsilon_{1}\cos\epsilon_{1},0)$. They form an orthogonal coordinate system.}
\label{fig:frame}
\end{figure*}

The detector coordinate system is chosen such that the three arms of the triangle detector keep fixed in the $x-z$ plane with the first spacecraft~(SC) located at $\vec{x}_{1}=(0,0,0)$, the second located at $\vec{x}_{2}=(0,0,L)$ and the third located at $ \vec{x}_{3}=(L\sin\gamma,0,L\cos\gamma)$, where $\gamma$ is the opening angle between the two arms of the interferometer and $L$ is the static length of the arm. The wave propagation direction is  
\[
\hat{k}=(\sin\theta_{1}\cos\epsilon_{1}, \sin\theta_{1}\sin\epsilon_{1},\cos\theta_{1}),
\]
and two orthogonal vectors $\hat{u}=(\cos\theta_{1}\cos\epsilon_{1}, \cos\theta_{1}\sin\epsilon_{1},-\sin\theta_{1})$, $\hat{v}=(-\sin\epsilon_{1}, \cos\epsilon_{1},0)$ can be chosen as the polarization basis vectors. The details of coordinate system is shown in Fig.~\ref{fig:frame}.

\subsection{Transfer functions of a single link}
Now we can calculate the frequency shift of laser caused by the motion of test masses and we will derive the one-link transfer functions in this section, which are essential for the discussions of transfer functions of TDI combinations.

Since the test masses will oscillate under the influence of ULBF, the time $T_{rs}$ that the laser need to propagate from sender SC ($\vec{x}_{s}$) to receiver SC ($\vec{x}_{r}$) will vary with time, 
\begin{equation}
    T_{rs} = L_{rs} + \delta t_{rs},
\end{equation}
where $L_{rs}$ is the static length of the arm in the absence of ULBF and $\delta t_{rs}$ is perturbation of one-way time due to the ULBF oscillation,
\begin{equation}\label{timevar}
    \delta t_{rs} = -\hat{n}_{rs} \cdot \left[\delta \vec{x}(t,\vec{x}_{r})-\delta \vec{x}(t-L,\vec{x}_{s})\right],
\end{equation}
where $\hat{n}_{rs}$ is the unit vector pointing from $\vec{x}_{r}$ to $\vec{x}_{s}$ and $t$ is the arrival time of laser recorded by the receiver. The laser phase $\phi_r$ received at $t$ equals the laser phase $\phi_s$ sent at $t-T_{rs}$,
\begin{equation}
    \phi_{r}(t) = \phi_{s}(t-T_{rs}) = 2\pi\nu_0(t-L_{rs}-\delta t_{rs}),
\end{equation}
where $\nu_0$ is the fiducial frequency of laser. The instantaneous frequency $\nu$ measured on the receiving spacecraft is related to $\phi_r$ through
\begin{eqnarray}
    \nu_{rs}(t) = \frac{1}{2\pi}\frac{d\phi_r}{dt}.
\end{eqnarray}
Thus, the relative frequency shift or fractional frequency fluctuation is
\begin{equation} \label{frevar}
    \frac{\delta \nu_{rs}}{\nu_0} = \frac{\nu_{rs} - \nu_{0}}{\nu_{0}} = -\frac{d\,\delta t_{rs}}{dt}.
\end{equation}
Combining Eq.~(\ref{timevar}) and Eq.~(\ref{frevar}), one-way Doppler shift can be expressed as 
\begin{equation}\label{frevar2}
    y_{rs}(t) \equiv \frac{\delta \nu_{rs}}{\nu_{0}} = \mu_{rs}\;[h(t,\vec{x_{r}})-h(t-L,\vec{x}_{s})] ,
\end{equation}
where $h(t,\vec{x})=m \mathcal{M}e^{im(t-v\hat{k}\cdot \vec{x})}$ and $\mu_{rs}$ is the geometric factor given by
\begin{equation}\label{mu}
    \mu_{rs} = \begin{cases}
        \hat{k}\cdot \hat{n}_{rs}\; & \;\; \text{for scalar field},\\
        \hat{e}_{A}\cdot \hat{n}_{rs}\; & \;\; \text{for vector field},\\
        \cfrac{\hat{n}^i_{rs}\hat{n}^j_{rs} {e}_{ij}(\hat{k},\psi)}{2(1+\hat{n}_{rs}\cdot \hat{k})}\; & \;\; \text{for gravitational wave},\\
    \end{cases}
\end{equation}
where we have included the response for gravitational wave for comparison, $e_{ij}(\hat{k},\psi)$ is the polarization tensor and $\psi$ is the polarization angle for gravitational wave. As can be seen from Eq.~(\ref{frevar2}), with the definition of $\mu_{rs}$, the one-way Doppler signals of scalar fields, vector fields and gravitational waves can be written in a concise and unified form. Follow from it, the TDI combinations, which are composed of time-shifted one-way Doppler signals, also have identical forms for scalar fields, vector fields and gravitational waves.

If the Doppler shift is recorded within a period of time $T$, the corresponding Fourier transform of $h(t)$ becomes
\begin{equation}\label{hF}
        h(t) = \frac{\sqrt{T}}{2\pi}\int_{0}^{T} \tilde{h}(\omega)e^{i\omega t} d\omega .
\end{equation}
Putting Eq.(\ref{hF}) in Eq.(\ref{frevar2}), we obtain
\begin{eqnarray}\label{signal}
    y_{rs}(t) = \mu_{rs} \; \frac{\sqrt{T}}{2\pi} \int_{0}^{T}  d\omega \; \tilde{h}(\omega) e^{i\omega t} \left[e^{-i(\vec{k}\cdot \vec{x}_{r})}-e^{-i(\tau +\vec{k}\cdot \vec{x}_{s})} \right],
\end{eqnarray}
where $\tau = \omega L\equiv 2\pi f L$ and $\tilde{h}(\omega)$ is the Fourier amplitude of $h(t)$. 
Then we can extract the Fourier transformation of $y_{rs}(t)$ from Eq.~(\ref{signal}):
\begin{equation}\label{sinam}
    \tilde{y}_{rs}(\omega) = \mu_{rs} \; \tilde{h}(\omega) \left[e^{-i(\vec{k}\cdot \vec{x}_{r})}-e^{-i(\tau +\vec{k}\cdot \vec{x}_{s})} \right].
\end{equation}
Here we have neglected the time-dependence of $\hat{n}_{rs}$ in $\mu_{rs}$, which follows the convention in the gravitational-wave literature and is compensated by the average of different directions.

The sky and polarization averaged transfer function $R(\omega)$ is defined as 
\begin{equation}
    R(\omega) = \overline{\left| \frac{\tilde{y}_{rs}(\omega)}{\tilde{h}(\omega)} \right|^{2}},
\end{equation}
where the overline denotes the average over all propagation directions and polarization angles. Explicitly, for scalar field, we have
\begin{equation} \label{transfer_s}
R^{s}(\omega)=\frac{1}{4\pi}
\int_{-1}^{1}d\cos\theta_{1} \int_{0}^{2\pi}d\epsilon_{1}
\left|\frac{\tilde{y}_{rs}(\omega,\theta_{1},\epsilon_{1})}{\tilde{h}(\omega)}\right|^{2}.
\end{equation}
For vector field, we have
\begin{equation} \label{transfer_v}
R^{v}(\omega)=\frac{1}{16\pi^{2}}\int_{-1}^{1}d\cos\theta_{1} \int_{0}^{2\pi}d\epsilon_{1}\int_{-1}^{1}d\cos\theta_{2} \int_{0}^{2\pi}d\epsilon_{2}\left|\frac{\tilde{y}_{rs}(\omega,\theta_{1},\epsilon_{1},\theta_{2},\epsilon_{2})}{\tilde{h}(\omega)}\right|^{2},
\end{equation}
where $(\theta_{2},\epsilon_{2})$ is the two-dimensional spherical coordinates of $\hat{e}_A$. For gravitational waves, we have
\begin{equation}
R^{GW}(\omega)=\frac{1}{8\pi^{2}}\int_{-1}^{1}d\cos\theta_{1} \int_{0}^{2\pi}d\epsilon_{1}\int_{0}^{2\pi}d\psi \left|\frac{\tilde{y}_{rs}(\omega,\theta_{1},\epsilon_{1},\psi)}{\tilde{h}(\omega)}\right|^{2}.
\end{equation}
We define the integration operations for later convenience,
\begin{align}
    I_s &\equiv \frac{1}{4\pi} \int_{-1}^{1}d\cos\theta_{1} \int_{0}^{2\pi}d\epsilon_{1} \cdots, \\
    I_v &\equiv \frac{1}{16\pi^{2}}\int_{-1}^{1}d\cos\theta_{1} \int_{0}^{2\pi}d\epsilon_{1}\int_{-1}^{1}d\cos\theta_{2} \int_{0}^{2\pi}d\epsilon_{2} \cdots, \\
    I_{GW}&\equiv \frac{1}{8\pi^{2}}\int_{-1}^{1}d\cos\theta_{1} \int_{0}^{2\pi}d\epsilon_{1}\int_{0}^{2\pi}d\psi \cdots .
\end{align}

\subsection{Transfer functions of TDI combinations}
As can be seen from Eq.~(\ref{signal}), the signal in one-link is a modulated Doppler shift with amplitude proportional to $\tilde{h}(\omega)$ whose typical magnitudes for astrophysical sources are of order $10^{-21}$. However, the current baseline for stabilized laser has a power spectral density as high as $30\;\text{Hz}/\sqrt{\text{Hz}}$ or $10^{-13}/\sqrt{\text{Hz}}$ in relative frequency fluctuation for laser with wavelength $\lambda=1064$~nm. Thus the laser frequency noise dominates in the one-link signals. In ground-based interferometer detectors, we do not need to worry about the laser noise in one link signals, since it will be suppressed significantly when the beams are combined at the photon detector due to the equal length of the two arms of Michelson interferometer. However, for space-borne interferometer detectors, the distances between pairs of spacecrafts evolve with time and can not keep equal with each other. Therefore the laser noise will not be cancelled by the simple Michelson configuration. To tackle this problem, TDI is used~\cite{Estabrook:2000ef, Tinto:2020fcc} for data postprocessing. The essential idea is to use the data of six links to synthesize virtual interferometric configurations that have almost equal light paths. It has been shown that the magnitude of residual laser noise in the synthesized data meets the requirement of sensitivity of gravitational waves detection.

In this section, we discuss the response of the TDI-1.5 configurations which apply to a non-equal arms, rigid but rotating triangle constellation. For the sake of simplicity and physical intuition, we focus on the equilateral-triangle configuration and ignore the motion of spacecrafts, since the final sensitivity curves do not depend on which generation TDI is. We perform the sky and polarization average of the transfer functions in both semi-analytical numerical integration and Monte Carlo integration with $39000$ source positions per frequency bin as a cross check of our results.

We follow the convention used in~\cite{Otto:2015erp} for TDI discussions. The opposite arm of spacecraft $i$ is denoted by $L_i$ (counter-clockwise) and $L_{i^{\prime}}$ (clockwise), respectively. $y(t)_{rs,ij^{\prime}} = y(t-L_i-L_{j^{\prime}})$ is the shorthand for time-delayed series.

\subsubsection{Michelson interferometry}
We start with equal-arm Michelson configuration $M(t)$, which can be synthesized from the four one-link signals,
\begin{equation}
    M(t) = [y_{12} + y_{21,3}] - [y_{13} + y_{31,2^\prime}].
\end{equation}
And in the static and equal armlength case, it reduces to
\begin{equation} \label{michelson}
    M(t) = [y_{12}(t) + y_{21}(t-L)] - [y_{13}(t) + y_{31}(t-L)].
\end{equation}
The corresponding Fourier transform $\tilde{M}(\omega)$ is given by 
\begin{equation} \label{michelsonf}
    \tilde{M}(\omega) = \tilde{h}(\omega) \; \left\{(\mu_{12}-\mu_{13})(1+e^{-i2\tau})e^{-i\vec{k}\cdot \vec{x}_{1}} - 2\left[ \mu_{12}e^{-i(\tau+\vec{k}\cdot \vec{x}_{2})} - \mu_{13}e^{-i(\tau+\vec{k}\cdot \vec{x}_{3})}\right ] \right\},
\end{equation}
where $\tau \equiv 2\pi f L$. After substituting Eq.~(\ref{michelsonf}) into Eqs.~(\ref{transfer_s}) and (\ref{transfer_v}), the sky and polarization averaged transfer functions $R^{s}_M$, $R^{v}_M$ of Michelson configuration can be expressed as 
\begin{eqnarray}\label{Sde}
        R^s_{M}(\omega) &=& I_{s} \left[ A_{1}^{2}(\theta_{1},\epsilon_{1},\omega) + A_{2}^{2}(\theta_{1},\epsilon_{1},\omega) \right] ,\\
        R^v_{M}(\omega) &=& I_{v} \left[ A_{1}^{2}(\theta_{1},\theta_{2},\epsilon_{1},\epsilon_{2},\omega) + A_{2}^{2}(\theta_{1},\theta_{2},\epsilon_{1},\epsilon_{2},\omega) \right],
\end{eqnarray}
and the expressions of $A_1$ and $A_2$ can be found in Appendix~\ref{sec:appendixA}.

 \subsubsection{Michelson combinations}
The Michelson $X$ combination in TDI-1.5 is given by
\begin{equation}
    X(t) = \left[ y_{13}+y_{31,2^\prime}+y_{12,22^\prime}+y_{21,322^\prime} \right] - \left[y_{12}+y_{21,3}+y_{13,3^\prime3}+y_{31,2^\prime3^\prime3} \right].
\end{equation}
For the static and equal armlength case,
\begin{eqnarray}\label{tdiX}
            X(t) &=& \left[ y_{13}(t)+y_{31}(t-L)+y_{12}(t-2L)+y_{21}(t-3L) \right] \nonumber \\
              &&{} - \left[y_{12}(t)+y_{21}(t-L)+y_{13}(t-2L)+y_{31}(t-3L) \right].
\end{eqnarray}
Michelson $Y$ and $Z$ combinations can be obtained by cyclic permutation of the indices $(1\rightarrow 2\rightarrow3\rightarrow 1)$.
The Fourier transform $\tilde{X}(\omega)$ is related to $\tilde{M}(\omega)$ through
\begin{equation} \label{XM}
    \tilde{X}(\omega) = -(1-e^{-i2\tau})\tilde{M}(\omega).
\end{equation}
This relation is a general consequence of the similarity between structures of Eq.~(\ref{michelson}) and Eq.~(\ref{tdiX}).
Using Eq.~(\ref{XM}), the averaged transfer functions $R^{s}$, $R^{v}$ of $X$ combination can be expressed as 
\begin{eqnarray}
        \label{Xs} R^s_{X}(\omega) &=& 4\sin^2\tau \; R^s_{M}(\omega), \\ 
        \label{Xv} R^v_{X}(\omega) &=& 4\sin^2\tau \; R^v_{M}(\omega).
\end{eqnarray}
After sky and polarization averaged, the transfer functions of $Y$, $Z$ combinations are the same as Eq.~(\ref{Xs}, \ref{Xv}).

In the first plot in Fig.~\ref{fig:tdi}, we show the transfer functions for Michelson combinations. We can see that in the low frequency range $ 2\times10^{-5} < f < 0.1$~Hz which corresponds to $v < \tau < 1$ ($v$ is the velocity of the field), $R^s_{X}, R^v_{X} \propto f^{6}$. However, when $f < 2\times10^{-5}$~Hz, namely, $\tau < v$, $R^s_{X}, R^v_{X} \propto f^{4}$. This is accord with the asymptotic behavior of GWs, $R^{GW}_{X} \propto f^{4}$. Thus, $f_c = v/2\pi L$ is the critical frequency at which the asymptotic behaviors change. This effect is due to the velocity of the scalar and vector fields. 

In the long wavelength limit, we can perform the sky average analytically and Eq.~(\ref{Xs}) reduces to 
\begin{equation} \label{Xslf}
    R^s_{X}(\omega) \simeq 16 \; \left[ \frac{4}{3}\sin^2\left(\frac{\gamma}{2}\right) \frac{\tau^6}{4}
                        + \frac{4}{15} \sin^2\gamma \cdot v^2\tau^4\right].
\end{equation}
For non-relativistic fields, the field velocity $v$ is a small quantity. The first term is dominant when $v < \tau$ and the second term become more important when $\tau < v$.

\subsubsection{Sagnac combinations: $\alpha, \beta, \gamma$}
The six-link Sagnac combinations $\alpha$, $\beta$, $\gamma$ are the generators of the space of laser noise free combinations. 
The expression of $\alpha$ combination is 
\begin{equation}
    \alpha(t) = \left[y_{13}+y_{32,2^\prime}+y_{21,1^\prime2^\prime} \right] - \left[ y_{12}+y_{23,3}+y_{31,13} \right],
\end{equation}
which in the static and equal armlength case reduces to 
\begin{equation}
    \alpha(t) = \left[ y_{13}(t)+y_{32}(t-L)+y_{21}(t-2L) \right] - \left[ y_{12}(t)+y_{23}(t-L)+y_{31}(t-2L) \right].
\end{equation}
$\beta$, $\gamma$ can be obtained by cyclic permutation of the indices (1,2,3).
The transfer function $R^{s}$, $R^{v}$ of $\alpha$ combination can be expressed as 
\begin{eqnarray} 
    \label{alphas} R^s_{\alpha}(\omega) &=& I_{s}[B_{1}^{2}(\theta_{1},\epsilon_{1},\omega) + B_{2}^{2}(\theta_{1},\epsilon_{1},\omega)], \\
    \label{alphav} R^v_{\alpha}(\omega) &=& I_{v}[B_{1}^{2}(\theta_{1},\theta_{2},\epsilon_{1},\epsilon_{2},\omega) + B_{2}^{2}(\theta_{1},\theta_{2},\epsilon_{1},\epsilon_{2},\omega)] ,
\end{eqnarray}
where the detail expressions of $B_1$ and $B_2$ are presented in the Appendix. The averaged transfer functions of $\beta$ and $\gamma$ are the same as Eq.~(\ref{alphas}, \ref{alphav}). 

The asymptotic behaviors of Sagnac combinations in the low frequency region are the same as the behaviors of Michelson combinations.
In Fig.~\ref{fig:tdi}, we can see that in the low frequency range $ 2\times10^{-5} < f < 0.1$~Hz, $R^s_{\alpha}, R^v_{\alpha} \propto f^{6}$. However, when $f < 2\times10^{-5}$~Hz, $R^s_{\alpha}, R^v_{\alpha} \propto f^{4}$, same as the asymptotic behavior of GWs, $R^{GW}_{\alpha} \propto f^{4}$.

\subsubsection{Fully symmetric Sagnac combination: $\zeta$}
Another important combination is the fully symmetric combination $\zeta$, given by
\begin{equation}
    \zeta(t) = y_{21,2^\prime}-y_{12,1}+y_{13,1^\prime}-y_{31,3}+y_{32,3^\prime}-y_{23,2}.
\end{equation}
For the static and equal armlength case, we have
\begin{equation}
    \zeta(t) = y_{21}(t-L)-y_{12}(t-L)+y_{13}(t-L)-y_{31}(t-L)+y_{32}(t-L)-y_{23}(t-L).
\end{equation}
The averaged transfer functions of $\zeta$ combination can be expressed as 
\begin{eqnarray}
    \label{rst} R^s_{\zeta}(\omega) &=& 4 \sin^2\left( \frac{\tau}{2} \right) \; I_{s} \left[C_{1}^{2}(\theta_{1},\epsilon_{1},\omega) + C_{2}^{2}(\theta_{1},\epsilon_{1},\omega) \right] ,\\
    \label{rvt} R^v_{\zeta}(\omega) &=& 4 \sin^2\left( \frac{\tau}{2} \right) \; I_{v} \left[C_{1}^{2}(\theta_{1},\theta_{2},\epsilon_{1},\epsilon_{2},\omega) + C_{2}^{2}(\theta_{1},\theta_{2},\epsilon_{1},\epsilon_{2},\omega) \right] .
\end{eqnarray}
The expressions of $C_1$ and $C_2$ can be found in the Appendix.

From Fig.~\ref{fig:tdi}, it is clear to see that in the low frequency limit $f<0.1$~Hz, $R^s_{\zeta}$, $R^{GW}_{\zeta} \propto f^{6}$ and $R^v_{\zeta} \propto f^{4}$. It is worth noting that the $f^6$ asymptotic behaviors of scalar field and GWs are true only when the detector array is equilateral triangle. If the array has unequal armlength, then $R^s_{\zeta}, R^{GW}_{\zeta}$ will be proportional to $f^{4}$ in the low frequency limit. At frequencies equal integers of the inverse of light travel time, the response of the detector is subtracted to zero due to the overall factor $\sin^2\left( \tau/2 \right)$ in Eq.~(\ref{rst}, \ref{rvt}). 

There are two interesting points here. The first is that $\zeta$ is not sensitive to the field velocity in the sense that there is only one asymptotic behavior of transfer function in the low frequency region and no critical frequency compared to Michelson combinations. The second is that the unexpected asymptotic behavior of vector field which is proportional to $f^{4}$ rather than $f^{6}$. 

To understand these phenomena, we derive the approximate expression of $\tilde{\zeta}(\omega)$ in the long wavelength limit,
\begin{equation} \label{zetalf}
    \tilde{\zeta}(\omega) \propto \left[ \frac{1}{3}(\mu_{12}\kappa_{12}
            +\mu_{23}\kappa_{23}+\mu_{31}\kappa_{31}) + \frac{2}{3}(\mu_{12}\kappa_{23}
            +\mu_{23}\kappa_{31}+\mu_{31}\kappa_{12})\right] (v\tau)^2 + O((v\tau)^3),
\end{equation}
where $\kappa_{ij} \equiv \hat{k}\cdot\hat{n}_{ij}$. It shows the field velocity $v$ and $\tau$ appear as a combination in Eq.~(\ref{zetalf}). Furthermore, for scalar field and the longitudinal polarization of vector field, $\mu_{ij} = \kappa_{ij}$, and the coefficient of $\tau^2$ cancels due to the geometrical relation $\kappa_{12}+\kappa_{23}+\kappa_{31} = 0$. Thus, the leading-order contribution is $O(\tau^3)$. However, for the transverse polarization of vector field, this cancellation doesn't happen and the leading order is $O(\tau^2)$. We further decompose the $R^v_{\zeta}$ into $R^{v,t}_{\zeta}$ for transverse mode and $R^{v,l}_{\zeta}$ for longitudinal mode. Following Eq.~(\ref{zetalf}), we find that $R^{v,l}_{\zeta} \propto f^6$ and $R^{v,t}_{\zeta} \propto f^4$ in the low frequency limit in the equal armlength case. Thus, transverse polarization dominates the asymptotic behavior of $R^v_{\zeta}$ in the low frequency region.

Besides serving as the noise monitor, $\zeta$ combination may have another advantage. If a monochromatic signal is detected, we need to distinguish ULBFs from galactic white dwarf binaries that can also produce nearly monochromatic gravitational waves. We can utilize the difference of transfer functions between $X$ and $\zeta$ combinations for ULBFs and GWs. Given a ULBF and a white dwarf binary which produce the same magnitude signals in the $X$ combination, the signal of ULBF in $\zeta$ combination may be overwhelmed by the noise due to the smaller value of transfer function while the signal of GWs in $\zeta$ combination will survive. This suggests that the ratios of signal-to-noise in two combinations are different in two cases. 

\begin{figure*}[!h]
    \centering
    \centerline{\includegraphics[scale=0.40]{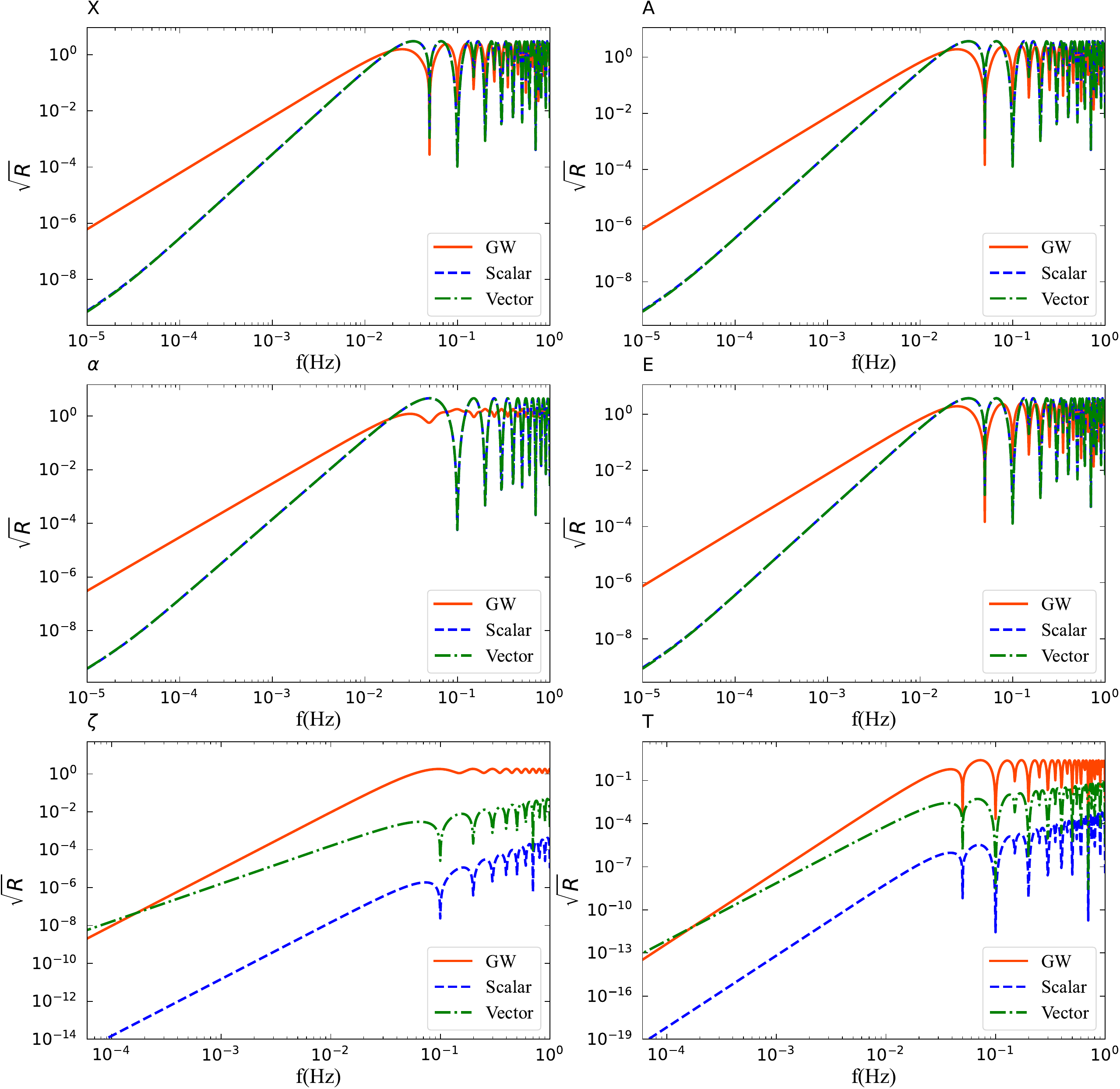}}
    \caption{The transfer functions of $X$, $\alpha$, $\zeta$, $A$, $E$ and $T$. We assume an equilateral triangle configuration with $L \simeq 3 \times 10^{9} $ m. For $X$, $\alpha$, $A$ and $E$, the transfer functions for scalar and vector almost overlap each other across the whole band.}
    \label{fig:tdi}
\end{figure*}

\subsubsection{Optimal combinations}
The Michelson combinations $X$, $Y$ and $Z$ can be further assembled into three optimal combinations, $A, E$ and $T$~\cite{Prince:2002hp},
\begin{eqnarray}
    A(t) &&= \frac{1}{\sqrt{2}} \left[ Z(t)-X(t) \right],\nonumber\\
    E(t) &&= \frac{1}{\sqrt{6}} \left[ X(t)-2Y(t)+Z(t) \right],\nonumber\\
    T(t) &&= \frac{1}{\sqrt{3}} \left[ X(t)+Y(t)+Z(t) \right].
\end{eqnarray}
The expressions of the averaged transfer functions of $A, E$, and $T$ can be found in the Appendix. As can be seen in Fig.~\ref{fig:tdi}, $R^{s}_A$, $R^{v}_A$ $\propto f^6$ in $ 2\times10^{-5} < f < 0.1$~Hz and $R^{s}_A$, $R^{v}_A$ $\propto f^4$ when $f < 2\times10^{-5}$~Hz (the values of $R^{s,v}_A$ and $R^{s,v}_E$ are equal across the band). In the equal armlength case, $R^{s}_T$, $R^{GW}_T$ $\propto f^{10}$ and $R^{v}_T$ $\propto f^8$ in the low frequency region. The different behavior of vector field is also due to the similar reason that the transverse polarization ($\propto f^8$) dominates the longitudinal polarization ($\propto f^{10}$) as explained in $\zeta$.

\subsection{Comparison of transfer functions of ULBFs with GWs} \label{subsec: velocity effect}
In the previous calculations, we adopted the assumption that $v \sim 10^{-3}$, which results in a different shape of transfer function compared to the case of gravitational waves, including the oscillating behavior in the high-frequency region and the trends in low-frequency region. Now we give a physical explanation for the difference. 

In previous sections, we obtain the displacement caused by bosonic fields in the non-relativistic case. In fact, we can directly apply the displacement formula of the detector under the influence of an ultralight scalar field to the relativistic case. This is because in the calculations we did not impose any assumption on the velocity of scalar field. Thus, the displacement formula derived earlier can still be used. But one should notice that, if the scalar field is relativistic, the approximation of $\omega\sim m$ is no longer applicable and should be replaced with $\omega\sim m/\sqrt{1-v^{2}}$. 

For simplicity, we will first compare the single-link responses and then discuss the transfer functions of TDI combinations.

\begin{figure} [!h]
\centerline{\includegraphics[scale=0.25]{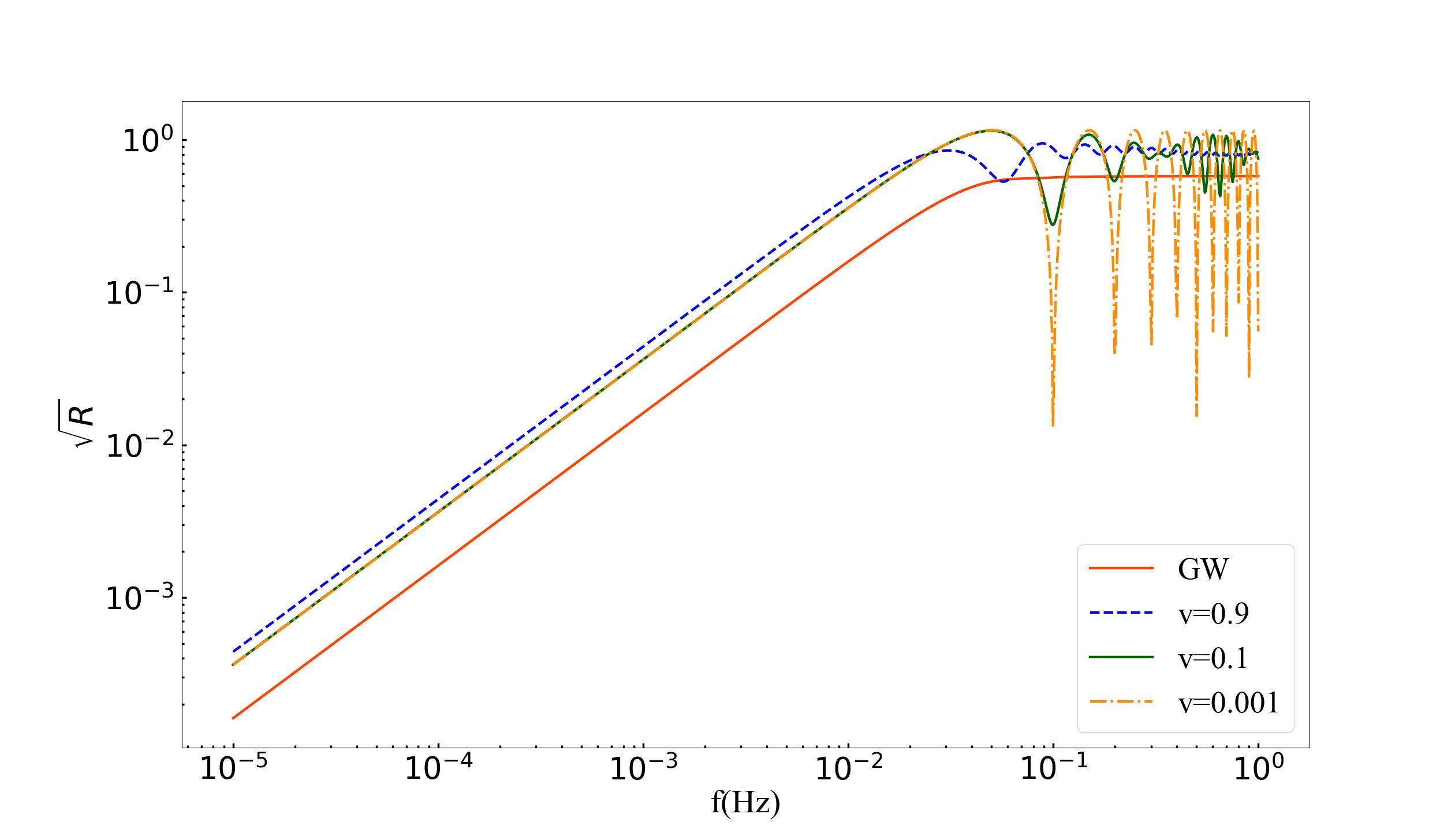}}
\caption{The single-link transfer function of scalar field. Here we use $L \simeq 3\times 10^9$~m and compare the transfer function of gravitational wave with that of scalar field with velocities $10^{-3}$, $0.1$ and $0.9$, respectively. }
\label{fig:relv}
\end{figure}

The Fourier amplitude of the single-link signal $y_{rs}(t)$ is given by
Eq.(\ref{sinam}) and the corresponding transfer function is 
\begin{equation}\label{tri}
    R^{s}_{single}(\omega) = I_{s}\mu_{rs}^{2} \left|e^{-i(\vec{k}\cdot \vec{x}_{r})}-e^{-i(\tau +\vec{k}\cdot \vec{x}_{s})} \right|^{2} = I_{s}\mu_{rs}^{2} \left[2-2\cos(\tau+L\vec{k}\cdot \hat{n}_{rs}) \right].
\end{equation}
For scalar field, $|\vec{k}|=\omega v$, different from $|\vec{k}|=\omega $ for gravitational waves. Without lose of generality, we use the arm $\hat{n}_{23}$ to calculate the single-link response function in the case of scalar field and gravitational waves. The result is shown in Fig.~\ref{fig:relv}. In the low-frequency region, the transfer function of the scalar field differs from that of the gravitational waves by a velocity-dependent factor. The leading-order term of the ratio in low frequency approximation is
\begin{equation}
    \frac{R^{s}_{single}(\omega)}{R^{GW}_{single}(\omega)} \approx
    \frac{I_{s}\mu_{rs}^{2}(1+v\hat{k}\cdot\hat{n}_{rs})^{2}}{I_{GW}\mu_{rs}^{2}(1+\hat{k}\cdot\hat{n}_{rs})^{2}}=5+3v^{2}.
\end{equation}
In the high frequency region, the velocity of the field $v$ affects the oscillation behavior. To be more specific, as $v$ increases, the second term related to the propagation direction in the trigonometric function in Eq.~(\ref{tri}) becomes non-negligible, which will lead to the change of the oscillation period. At the same time, averaging the entire sky will result in a decrease in the oscillation amplitude, since while cancellation appear in some directions at a fixed frequency, it will not happen in other directions.

\begin{figure} [!h]
\centering
\centerline{\includegraphics[scale=0.20]{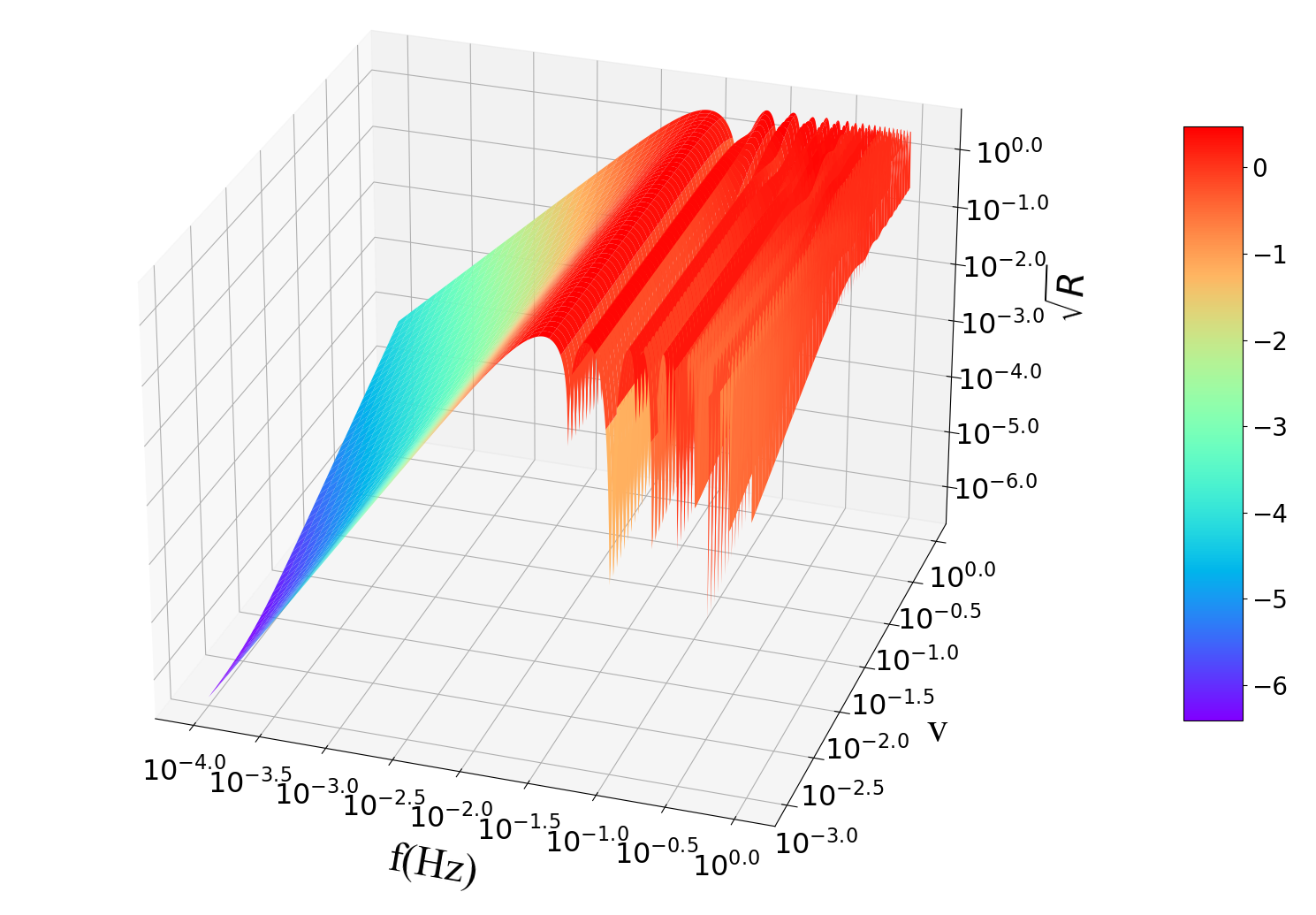}}
\caption{The transfer function of $X$ combination for scalar field, using velocity of scalar field and frequency as independent variables. We use the arm length $L \simeq 3\times 10^9$~m.
\label{fig:3d}}
\end{figure}

Now we consider the TDI combinations. The differences between transfer functions are similar to the single-link case.
We illustrate with the transfer function of $X$ combination with respect to the frequency and the velocity in Fig.~\ref{fig:3d}. 

From the result, we find that with the increasing velocity, the overall trend of the transfer function and the oscillation behavior of the high frequency part are closer to the behavior of gravitational wave. When the velocity of the scalar field approaches the speed of light, there is only a slight difference between the two cases, which is caused by $\mu_{rs}$.
 
\section{Sensitivity curves}\label{sec:sensitivity}

\begin{figure}[!h]
    \centering
    \centerline{\includegraphics[scale=0.50]{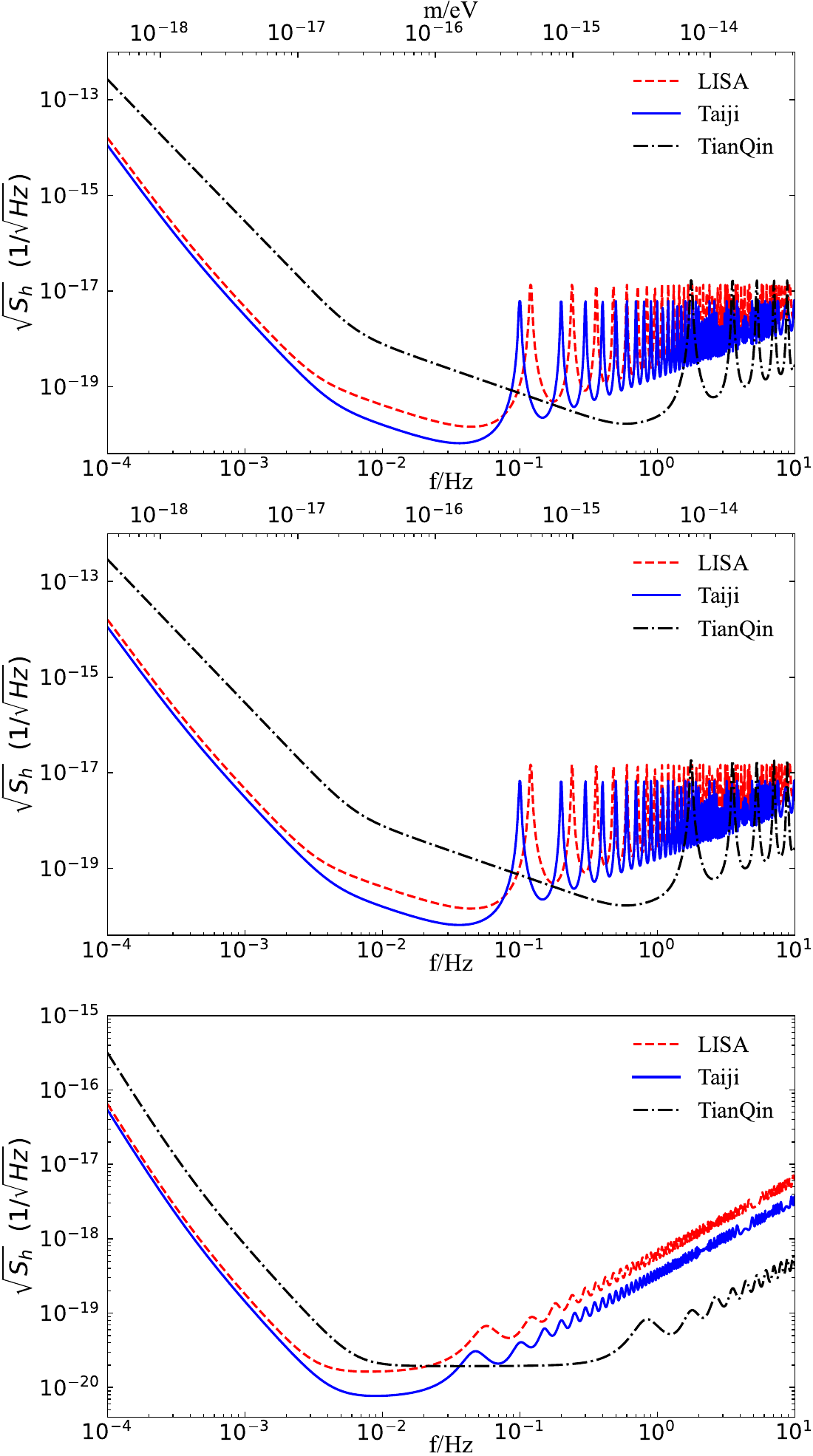}}
    \caption{The sensitivity curves of LISA, Taiji and TianQin to scalar field (top), vector field (middle) and gravitational wave (bottom). We adopt $L_{\textrm{LISA}}=2.5\times10^9$~m, $L_{\textrm{Taiji}}=3\times10^9$~m and $L_{\textrm{TianQin}}=1.7\times10^8$~m in the calculation.}
    \label{fig:ltt}
\end{figure}

The sensitivity of individual TDI combination $O$ is defined by \cite{babak2021lisa}
\begin{equation}\label{sen}
    S_{O}(f) = \frac{N_{O}(f)}{R_{O}(f)},
\end{equation}
where $N_O(f)$ is the one-sided noise power spectral density (PSD) of combination $O$ and $R_O(f)$ is the averaged transfer function of $O$.

After suppression of the laser noise below the requirement, the two secondary noises determine the performance of detector. The first one is related to optical metrology system (oms) noise, which dominates at high frequency. The second one is the test mass acceleration noise (acc) that dominates at low frequency. The magnitudes of oms and acc noises for LISA~\cite{LISA:2017pwj}, Taiji~\cite{Hu:2017mde}, TianQin~\cite{TianQin:2015yph} are
\begin{eqnarray}
        \label{oms} S_{oms}\left( f\right) &=& \left( s_{oms}\;\frac{2\pi f}{c} \right)^{2} \left[ 1 + \left(\frac{2 \times 10^{-3}~\textrm{Hz}}{f} \right)^{4} \right] \; \frac{1}{\textrm{Hz}}, \\
        \label{acc} S_{acc}\left( f\right) &=& \left( \frac{s_{acc}}{2\pi fc} \right)^{2} \left[ 1+\left(\frac{0.4 \times 10^{-3}~\textrm{Hz}}{f} \right)^{2} \right] \; \left[ 1+ \left(\frac{f}{8 \times 10^{-3}~\textrm{Hz}} \right)^{4} \right] \; \frac{1}{\textrm{Hz}},
\end{eqnarray}
where we use the following noise parameters,
\begin{eqnarray}\label{eq:omsacc}
   \textrm{LISA} &:& ~ s_{oms} = 15\times10^{-12}~\textrm{m}, s_{acc} = 3\times10^{-15} ~\textrm{m}/\textrm{s}^2, \nonumber \\
   \textrm{Taiji}&:& ~ s_{oms} = 8\times10^{-12}~\textrm{m}, s_{acc} = 3\times10^{-15} ~\textrm{m}/\textrm{s}^2, \nonumber \\
   \textrm{TianQin}&:& ~ s_{oms} = 1\times10^{-12}~\textrm{m}, s_{acc} = 1\times10^{-15} ~\textrm{m}/\textrm{s}^2.
\end{eqnarray}
We adopt the same frequency dependent factors in Eqs.~(\ref{oms}, \ref{acc}) for a straightforward comparison of the sensitivity of the three detectors. 

Since we perform TDI algorithm with the one-link output, the remaining noises in the output of TDI are also the linear combinations of the time shifted noises in one-link output. We derived the PSD for the TDI-1.5 combinations~(PSD of other TDI combinations can be found in Appendix~\ref{sec:appendixB}):
\begin{eqnarray}
    N_{X} &=& 16\sin^{2}(\tau) \left\{ \left[ 3+\cos(2\tau) \right]S_{acc} + S_{oms} \right\}, \\
    N_{A} &=& N_{E} = 8\sin^{2}(\tau)\; \left\{ 2\left[3+2\cos\tau+\cos(2\tau)\right]\;S_{acc} + \left(2+\cos\tau\right)\;S_{oms} \right\}, \\
    N_{T} &=& 32\sin^{2}\left(\frac{\tau}{2}\right)\sin^{2}(\tau) \; \left[4\sin^{2}\left(\frac{\tau}{2}\right)\;S_{acc} + S_{oms} \right].
\end{eqnarray}

With the averaged transfer function and PSD of the TDI combinations at hand, we can calculate  sensitivity curves. In Fig.~\ref{fig:ltt}, we plot the sensitivity  of Michelson combination $S_{X}$ of the three detectors to the ULBFs. 

\begin{figure}[h]
    \centering
    \centerline{\includegraphics[scale=0.25]{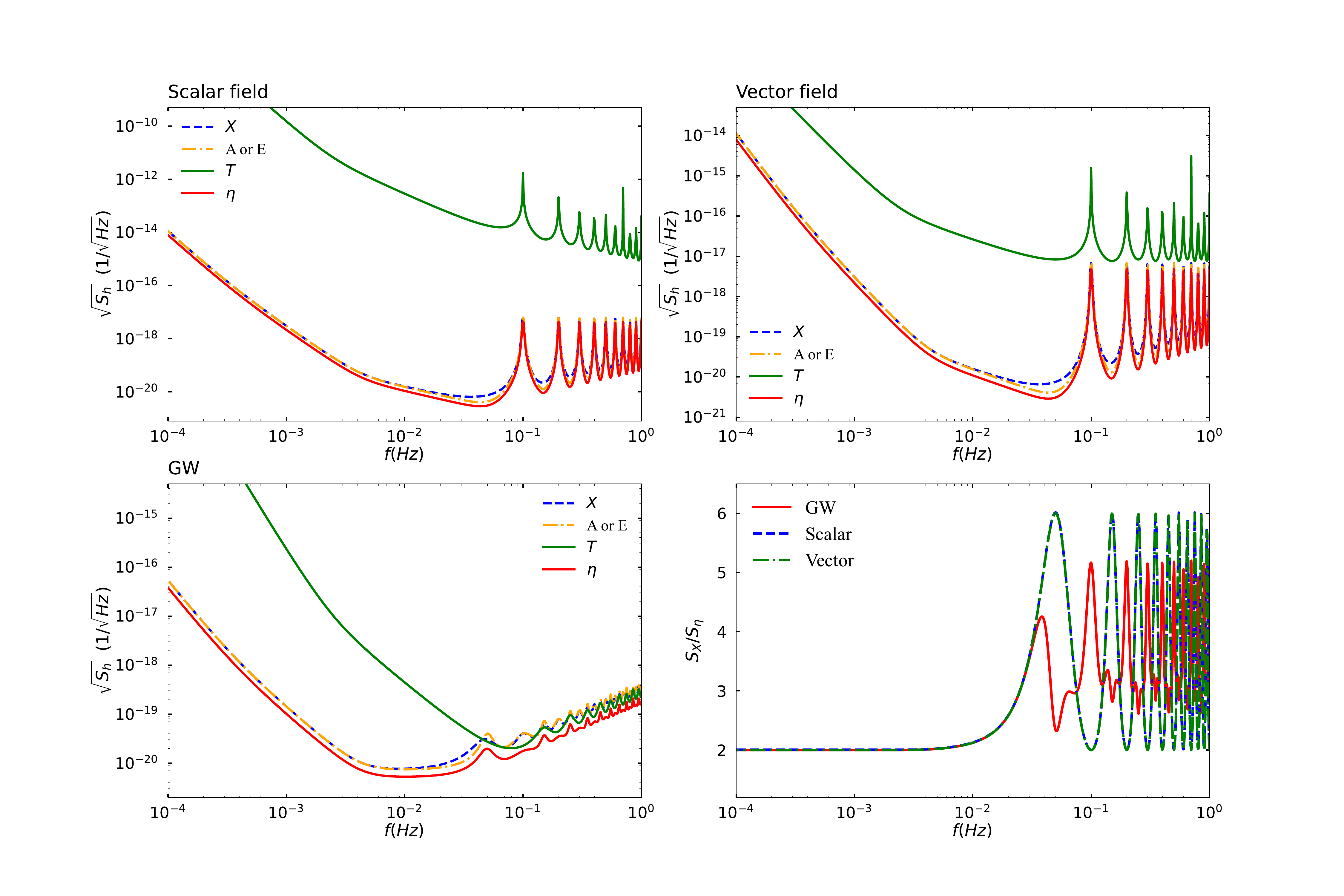}}
    \caption{The optimal sensitivity $S_{\eta}$ and the ratios of $S_{X}$ to $S_{\eta}$ for scalar field, vector field and gravitational wave.}
    \label{fig:optimal}
\end{figure}

The optimal sensitivity $S_{\eta}$ is defined by 
\begin{equation}
    \frac{1}{S_{\eta}} = \frac{1}{S_A}+\frac{1}{S_E}+\frac{1}{S_T} .
\end{equation} 
In Fig.~\ref{fig:optimal} we compare the optimal sensitivity with $X$ combination sensitivity for Taiji. For gravitational waves, the ratio between $S_{\eta}$ and $S_{X}$ oscillates with an average value $3$ in the high frequency part. However, for scalar and vector fields, the ratio oscillates above $2$ in the high frequency part. This difference comes from the relative magnitudes between $S_A$, $S_E$ and $S_T$ in the high frequency band in the two cases. As shown in Fig.~\ref{fig:optimal}, for gravitational waves, although in the long wavelength region the sensitivity of $A$ and $E$ are much better than $T$, $T$ will have comparable sensitivity in the high frequency region. For scalar and vector fields, $T$ has much worse sensitivity than $A$ and $E$ across the whole band, thus will not contribute to $S_{\eta}$ significantly.

\section{Illustrations of Constraints on Dark Matter}\label{sec:constraint}

\begin{figure}[!h]
\includegraphics[scale=0.25]{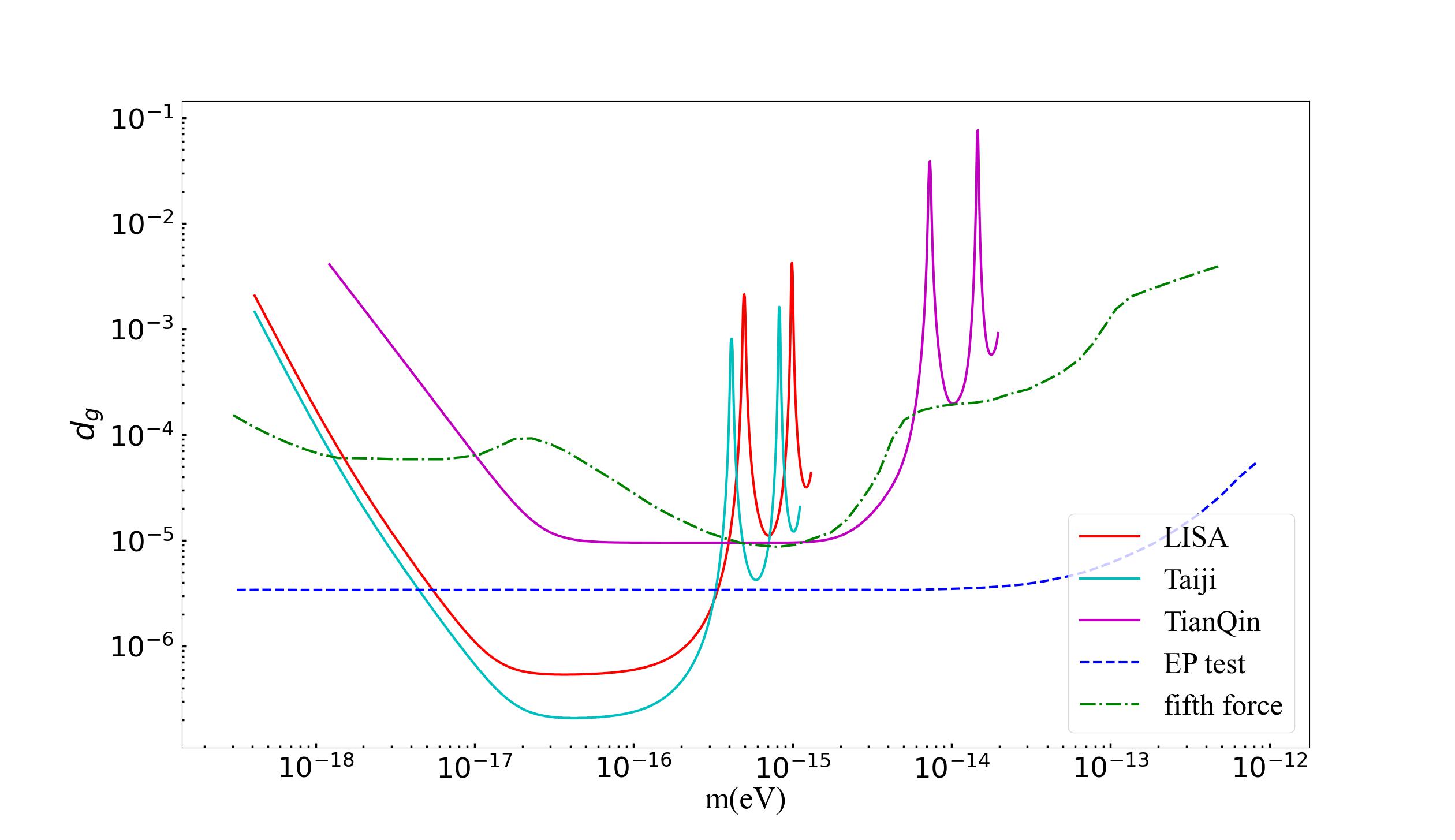}
\caption{The constraints on the coupling constant $d_{g}$ given by the future space-based gravitational-wave detectors: LISA, Taiji and TianQin. The observation time $T = 1 $ year. For comparison, the constraints given by the fifth-force experiment~\cite{doi:10.1146/annurev.nucl.53.041002.110503} and the the tests of the equivalence principle (EP tests) are also shown. The constraints of EP tests shown here are given by the experiment from the Eöt-Wash group~\cite{Wagner_2012} and MICROSCOPE \cite{PhysRevLett.119.231101}.  Regions above the lines have been excluded. We consider that the coupling constants other than $d_{g}$ are zero here.}
\label{fig:dg}
\end{figure}
\begin{figure}[!h]
  \centering
    \includegraphics[scale=0.23]{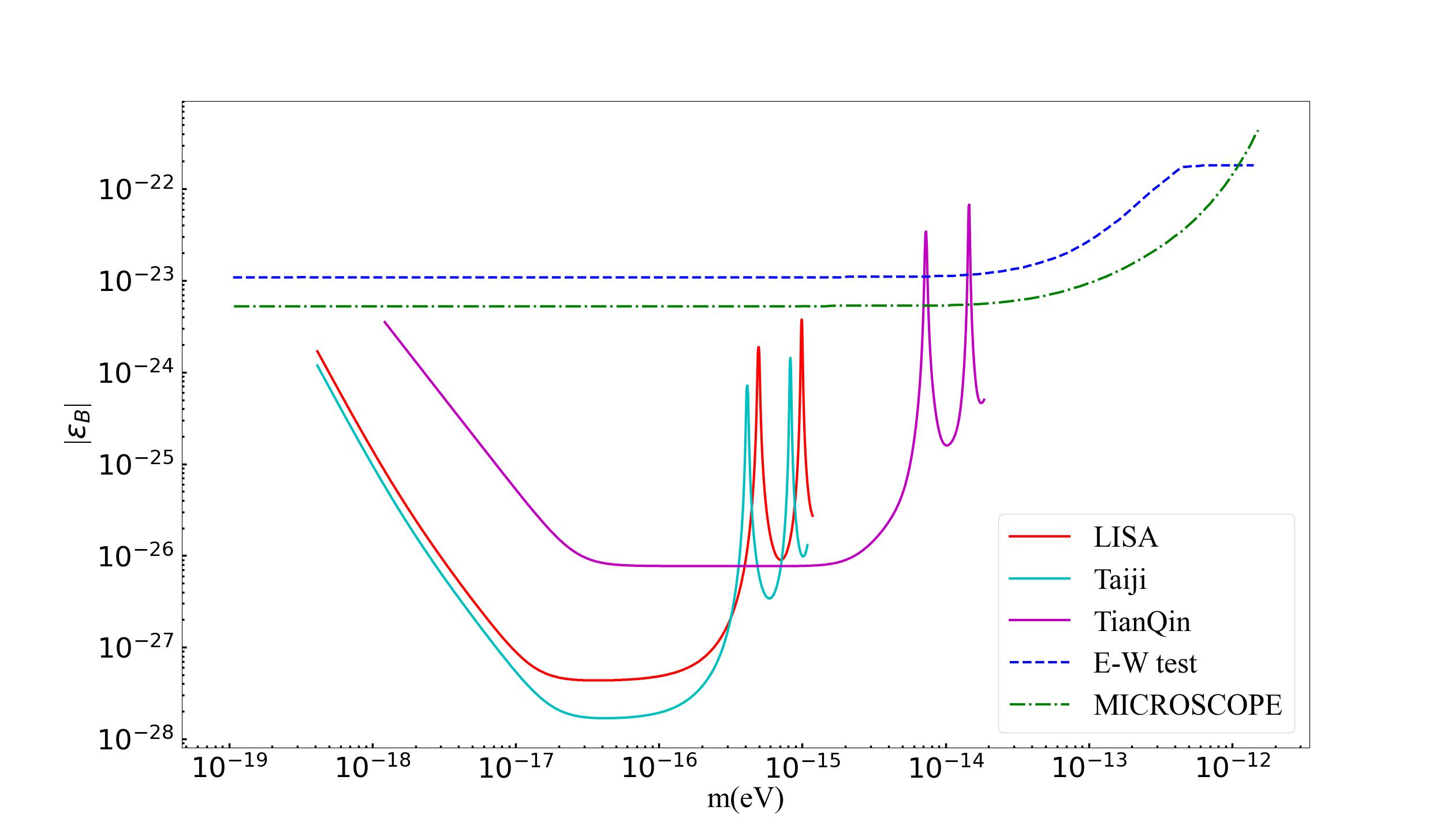}
    \includegraphics[scale=0.23]{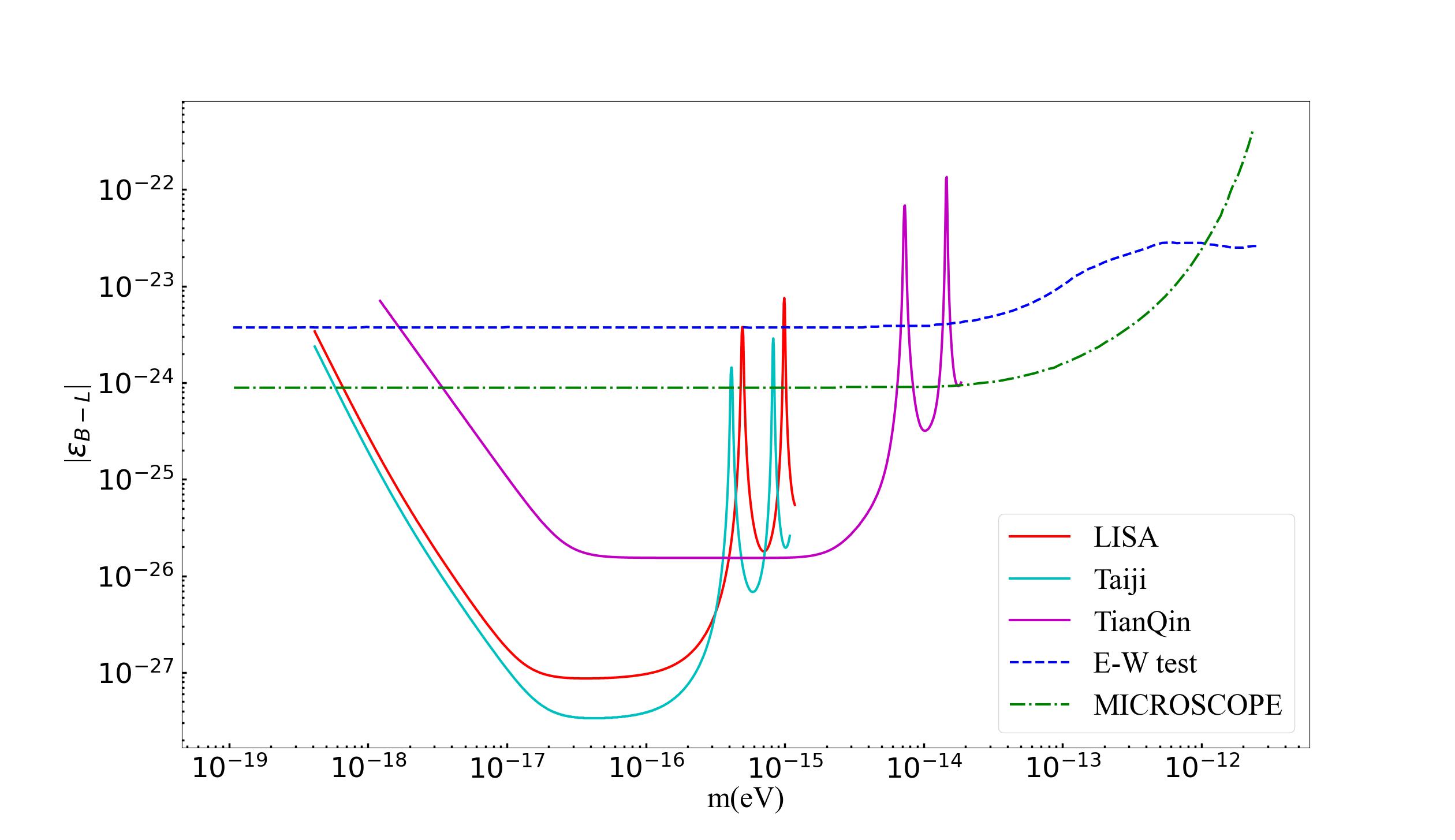}
  \caption{The constraints on the coupling constants $\epsilon_{B}$ and $\epsilon_{B-L}$ projected by the future space-based gravitational-wave detectors: LISA, Taiji and TianQin. The observation time $T = 1 $ year. For comparison, the constraints given by the tests of the equivalence principle (EP tests) are also shown. The constraints of EP tests shown here are given by the experiment from the Eöt-Wash group \cite{PhysRevLett.100.041101, Wagner_2012} and MICROSCOPE~\cite{PhysRevLett.120.141101}. Regions above the lines are excluded.}
  \label{fig:convec}
\end{figure}

In this section, we estimate the future constraints on dark matter (DM) models based on the results in the previous sections. 

For scalar field, we consider the following interactions as an illustration~\cite{PhysRevD.82.084033, Pospelov:2010ye},
\begin{equation}
    \mathcal{L}_{\phi-SM}=\kappa \phi \left[-\frac{d_{g}\beta_{3}}{2g_{3}}F^{A}_{\mu \nu}F^{A \mu \nu}-\sum_{i=u,d}(d_{m_{i}}+\gamma_{m_{i}}d_{g})m_{i}\bar{\psi}_{i}\psi_{i}\right],
\end{equation}
where $\kappa \equiv \sqrt{4\pi}/M_{P}$; $F^{A}_{\mu \nu}$ are the gluon field strength tensors; $g_{3}$ is the $SU(3)$ gauge coupling; and $\beta_{3}$ refers to the QCD beta functions; $\gamma_{m_{i}}$ are the anomalous dimensions of the $u$ and $d$ quarks; $d_{g}$ is the couplings to the gluonic field term, and $d_{m_{u}}$ and $d_{m_{d}}$ are the couplings to the quark mass terms. It is more convenient to define 
\begin{equation} \label{dm amplitude}
    d_{\hat{m}} \equiv \frac{d_{m_{d}}m_{d}+d_{m_{u}}m_{u}}{m_{d}+m_{u}}, \qquad  d_{\delta m}\equiv \frac{d_{m_{d}}m_{d}-d_{m_{u}}m_{u}}{m_{d}-m_{u}}.
\end{equation}
Since the nucleon mass is mostly determined by the QCD energy scale, $\alpha(\phi)$ in Eq.~(\ref{eq:alpha}) is approximately given by
\begin{equation}
    \alpha(\phi)\approx d_{g}^{*} \approx d_{g} + 0.093(d_{\hat{m}}-d_{g}).
\end{equation}

Considering that scalar or vector fields make up all the dark matter components, we have
\begin{equation}
    \phi_{\vec{k}} = \frac{\sqrt{2\rho_{\textrm{DM}}}}{m_{\phi}}, \qquad
    |\vec{A}| = \frac{\sqrt{2\rho_{\textrm{DM}}}}{m_{A}},
    \qquad
    v\sim v_{\textrm{DM}}\sim10^{-3},
\end{equation}
where $\rho_{\textrm{DM}} = 0.3$~GeV$/$cm$^{3}$ is an estimate of the local density of DM, and $v_{\textrm{DM}}$ is the velocity dispersion of the dark matter in our Galaxy.

The minimum detectable Fourier amplitude is 
\begin{equation}\label{sen2}
    \left|\tilde{h}(m_{\phi}) \right| =m_{\phi}\mathcal{M}_{s} \sqrt{T}
    =\frac{d_{g}^{*} \kappa \phi_{\vec{k}}|\Vec{k}|\sqrt{T}}{m_{\phi}} = \sqrt{S_{O}},
\end{equation}
where $T$ refers to the duration of observation time. We assume that the duration is shorter than the coherence time of the field $~4\times 10^{-16}\textrm{eV}/m_\phi \times 10^7$ s, so that the approximation of monochromatic plane wave is satisfied. Moreover, local DM should be treated as a superposition of plane waves with different propagation directions and velocities and a $\mathcal{O}(3)$ factor has been found in the estimation of laboratory experimental searches~\cite{Centers:2019dyn}. Here, we just use the monochromatic plane wave for illustration and more dedicated study on the stochastic DM will be conducted in the coming work.

Using the standard sensitivity curve of $X$ combination, we express the constrain $d_{g}^{*}$ with respect to $m_{\phi}$,
\begin{equation}\label{dgex}
    d_{g}^{*} = \frac{m_{\phi}}{\kappa v_{\textrm{DM}} }\sqrt{\frac{S_{X}}{2\rho_{\textrm{DM}}T}}.
\end{equation}
The constraint curves of coupling constant $d_{g}^{*}(m_{\phi})$ given by space-based detectors are plotted in Fig.~\ref{fig:dg}, assuming $d_m =0$ and $d_{g}^{*} \approx 0.9d_{g}$. Relevant constraints from other experiments, including the fifth-force experiment~\cite{doi:10.1146/annurev.nucl.53.041002.110503} and the tests of equivalence principle (EP tests)~\cite{Wagner_2012, PhysRevLett.119.231101}, are also shown. All detectors will provide more stringent constraints than the fifth-force experiments in some parameter space. The constraint given by TianQin can reach $9.5\times10^{-6}$, which approaches the lower limit of the fifth-force experiment. Compared to the EP tests, LISA and Taiji can reach $5.38\times 10^{-7}$ and $2.07\times 10^{-7}$ at $4.28\times 10^{-17}$ eV and $4.44\times 10^{-17}$ eV, respectively, and the improvement factors are 6 and 16, respectively. 
In the low mass region, Eq.~(\ref{dgex}) has the approximate expression, $d_{g}^{*} \propto \tau^{-2}\sqrt{S_{acc}}$, i.e.~the sensitivity is inversely proportional to the square of the arm length. Therefore, LISA and Taiji will have much better sensitivity of $d_{g}^{*}$ compared to TianQin at low-mass region.

Similarly, the Fourier amplitude of the vector field $\tilde{h}(\omega)$ is given by 
\begin{equation}\label{ho}
    |\tilde{h}(m_{A})| =m_{A}\mathcal{M}_{v} \sqrt{T}
    =\frac{ \epsilon_{D} e q_{D,j}|\vec{A}|\sqrt{T}}{M_{j}} = \sqrt{S_{O}}.
\end{equation}
Using the sensitivity curve of $X$ combination for vector fields, we have 
\begin{equation}
    \epsilon_{D} = \frac{m_{A}M_{j}}{e q_{D,j}}\sqrt{\frac{{S}_{X}}{2\rho_{\textrm{DM}}T}}.
\end{equation}
For a $U(1)_{B}$ gauge boson, $q_{B}/M=(A/\mu)/ m_{n}\approx 1/m_{n}$ is
almost identical among different materials, where $A$ is
the mass number, $\mu$ is the atomic mass in atomic units and $m_{n}$ refers to the mass of neutron. But for a $U(1)_{B-L}$ gauge boson, $q_{B-L}/M=[(A-Z)/\mu]/m_{n}\approx 0.5/m_{n}$, and the neutron ratio $(A-Z)/\mu$ would
be different between materials. Here we use $A/\mu=1$ for a $U(1)_{B}$ gauge boson and $(A-Z)/\mu=0.5$ for a $U(1)_{B-L}$ gauge boson. 

The constraints on coupling constants $|\epsilon_{B}|(m)$ and $|\epsilon_{B-L}|(m)$ given by space-based detectors are plotted in Fig.~\ref{fig:convec}. As a comparison, the results of Eöt-Wash experiment and MICROSCOPE are also shown~\footnote{The results of MICROSCOPE shown here are from the first results of the experiment~\cite{PhysRevLett.119.231101}. According to the latest results~\cite{PhysRevLett.129.121102}, the constraints would be improved by a factor of $\sim 2$. Private communication with P. Fayet.}.
Compared to EP test, gravitational-wave detectors can provide better constraints on coupling constants $|\epsilon_{B}|$ and $|\epsilon_{B-L}|$ in some parameter space. For $|\epsilon_{B}|$, the best constraints projected by LISA, Taiji and TianQin in the sensitive range are improved by $\sim 12000$, $\sim 31000$ and $\sim 680$ respectively. And for $|\epsilon_{B-L}|$, the improvement factors are $\sim 1000$, $\sim 2600$ and $\sim 60$. As we can see, LISA and Taiji will have much better sensitivity of $|\epsilon_{B}|$ and $|\epsilon_{B-L}|$ compared to TianQin at low-mass region because of the longer arm length.

\section{Summary and Conclusions}\label{sec:summary}

Ultralight bosonic fields (ULBFs) are well-motivated and predicted in many physical and cosmological theories. When having tiny couplings to standard model particles, they may be probed by precision gravitational-wave interferometers. We investigate the sensitivity of space-based gravitational-waves detectors to ULBFs systematically. We obtain the sky and polarization averaged transfer functions of various time-delay interferometry (TDI) combinations and discuss their asymptotic behaviors in the long wavelength region. We find that the velocity of field will affect the behaviors of transfer function, more significantly at high mass region. For TDI combinations like Michelson, Sagnac and optimal $A$, $E$, there exists a critical frequency $f_c = v/2\pi L$ determined by the field velocity $v$ and arm length $L$, below which transfer functions $R^{s,v} \propto f^4$ and are similar to gravitational waves; however, for $f_c < f < 0.1$~Hz, $R^{s,v}$ are proportional to $f^6$. For combinations like $\zeta$ and $T$, although there is no critical frequency, the magnitudes of transfer functions for scalar and vector fields are much smaller than that for gravitational waves. The physical reason is that, for the same frequency these non-relativistic waves will have much longer wavelength compared to gravitational waves, and $\zeta$ and $T$ are not sensitive to the common displacements of the spacecrafts due to their symmetric structure. 

A direct application of our result is to identify ULBFs as dark matter (DM). We estimate the ability of detectors to search DM by calculating the sensitivity curves of future space-based interferometers, which can improve the current limits by several orders of magnitude, see Figs.~\ref{fig:dg} and \ref{fig:convec}. We should note that it is more realistic and appropriate to treat the local DM as a stochastic waves background rather than monochromatic waves~\cite{Centers:2019dyn, PhysRevA.97.042506, PhysRevD.97.123006, Kim:2023pkx}. Therefore, the precise sensitivity of detectors to the local DM need a more dedicated study. We will leave the stochastic effect of the fields to future work.

\section*{acknowledgement}
Y.T is supported by National Key Research and Development Program of China (Grant No.2021YFC2201901), and National Natural Science Foundation of China under Grants No.12147103 and 11851302. Y.L.W. is supported by the National Key Research and Development Program of China under Grant No.2020YFC2201501, and NSFC under Grants No.~11690022, No.~11747601, No.~12147103, and the Strategic Priority Research Program of the Chinese Academy of Sciences under Grant No. XDB23030100.

\appendix

\section{Analytic expressions of the transfer functions} \label{sec:appendixA} 
Explicit forms of some functions appearing in the transfer functions of Michelson and Sagnac combinations are listed below.
\begin{eqnarray*}
        A_{1} =&&
        (\mu_{12}-\mu_{13})\left[ 1+\cos(2\tau) \right] -2[\mu_{12}\cos(\tau+\vec{k}\cdot \vec{x}_{2})-\mu_{13}\cos(\tau+\vec{k}\cdot \vec{x}_{3})],\\
        A_{2} =&& -
        (\mu_{12}-\mu_{13})\sin(2\tau)+2[\mu_{12}\sin(\tau+\vec{k}\cdot \vec{x}_{2})-\mu_{13}\sin(\tau+\vec{k}\cdot \vec{x}_{3})],
\end{eqnarray*}
\begin{eqnarray*}
        B_{1} = &&
        \mu_{21}[1-\cos3\tau-\cos(\vec{k}\cdot \vec{x}_{2}+\tau)-\cos(\vec{k}\cdot \vec{x}_{2}+2\tau)]+
        \mu_{13}[1-\cos3\tau-\cos(\vec{k}\cdot \vec{x}_{3}+\tau)
        -\nonumber\\&&
        \cos(\vec{k}\cdot \vec{x}_{3}+2\tau)]+\mu_{32}[\cos(\vec{k}\cdot \vec{x}_{2}+\tau)+\cos(\vec{k}\cdot \vec{x}_{3}+\tau)-\cos(\vec{k}\cdot \vec{x}_{2}+2\tau)-\cos(\vec{k}\cdot \vec{x}_{3}+2\tau)],\\
        B_{2} = &&
        \mu_{21}[\sin3\tau+\sin(\vec{k}\cdot \vec{x}_{2}+\tau)+\sin(\vec{k}\cdot \vec{x}_{2}+2\tau)]+\mu_{13}[\sin3\tau+\sin(\vec{k}\cdot \vec{x}_{3}+\tau)
        +\sin(\vec{k}\cdot \vec{x}_{3}+2\tau)]\nonumber\\&&
        +\mu_{32}[-\sin(\vec{k}\cdot \vec{x}_{2}+\tau)-\sin(\vec{k}\cdot \vec{x}_{3}+\tau)+\sin(\vec{k}\cdot \vec{x}_{2}+2\tau)+\sin(\vec{k}\cdot \vec{x}_{3}+2\tau)],
\end{eqnarray*}
\begin{eqnarray*}
         C_{1} =&&
         \mu_{21}(1+\cos(\vec{k}\cdot \vec{x}_{2}))+\mu_{13}(1+\cos(\vec{k}\cdot \vec{x}_{3}))+\mu_{32}[\cos(\vec{k}\cdot \vec{x}_{2})+\cos(\vec{k}\cdot \vec{x}_{3})],\\
        C_{2} =&& \mu_{21}\sin(\vec{k}\cdot \vec{x}_{2})+\mu_{13}\sin(\vec{k}\cdot \vec{x}_{3})+\mu_{32}[\sin(\vec{k}\cdot \vec{x}_{2})+\sin(\vec{k}\cdot \vec{x}_{3})].
\end{eqnarray*}
The transfer function $R_{s}$, $R_{v}$ for various optimal combinations can be expressed as 
\begin{eqnarray}
    R_{A}^s(\omega) &=& I_{s}(1-\cos 2\tau)[D_{1}^{2}(\theta_{1},\epsilon,\omega) + D_{2}^{2}(\theta_{1},\epsilon,\omega)],\\
    R_{A}^v(\omega)&=& I_{v}(1-\cos 2\tau)[D_{1}^{2}(\theta_{1},\theta_{2},\epsilon_{1},\epsilon_{2},\omega) + D_{2}^{2}(\theta_{1},\theta_{2},\epsilon_{1},\epsilon_{2},\omega)],
\end{eqnarray}
\begin{eqnarray}
     R_{E}^s(\omega)&=& I_{s}\frac{1}{3}(1-\cos 2\tau)[D_{3}^{2}(\theta_{1},\epsilon,\omega) + D_{4}^{2}(\theta_{1},\epsilon,\omega)] ,\\
    R_{E}^v(\omega)&=& I_{v}\frac{1}{3}(1-\cos 2\tau)[D_{3}^{2}(\theta_{1},\theta_{2},\epsilon_{1},\epsilon_{2},\omega) + D_{4}^{2}(\theta_{1},\theta_{2},\epsilon_{1},\epsilon_{2},\omega)],   
\end{eqnarray}
\begin{eqnarray}
    R_{T}^s(\omega)&=& I_{s}\frac{2}{3}(1-\cos 2\tau)[D_{5}^{2}(\theta_{1},\epsilon,\omega) + D_{6}^{2}(\theta_{1},\epsilon,\omega)],\\
    R_{T}^{v}(\omega)&=& I_{v}\frac{2}{3}(1-\cos 2\tau)[D_{5}^{2}(\theta_{1},\theta_{2},\epsilon_{1},\epsilon_{2},\omega) + D_{6}^{2}(\theta_{1},\theta_{2},\epsilon_{1},\epsilon_{2},\omega)],
\end{eqnarray}
where we have defined the following functions,
\begin{eqnarray*}
    D_{1} =&& -(\mu_{32}+\mu_{12})\left[ 1+\cos(2\tau) \right]+2\mu_{32}\cos(\tau+\vec{k}\cdot \vec{x}_{2}+\vec{k}\cdot \vec{x}_{3})+2\mu_{12}\cos(\tau+\vec{k}\cdot \vec{x}_{1}+\vec{k}\cdot \vec{x}_{2}),\\
    D_{2} =&&  (\mu_{32}+\mu_{12})\sin(2\tau)-2\mu_{32}\sin(\tau+\vec{k}\cdot \vec{x}_{2}+\vec{k}\cdot \vec{x}_{3})-2\mu_{12}\sin(\tau+\vec{k}\cdot \vec{x}_{1}+\vec{k}\cdot \vec{x}_{2}),\\
    D_{3} =&& -(2\mu_{13}+\mu_{23}-\mu_{21})\left[ 1+\cos(2\tau) \right]+4\mu_{13}\cos(\tau+\vec{k}\cdot \vec{x}_{1}+\vec{k}\cdot \vec{x}_{3}) \nonumber\\
        &&{}+2\mu_{23}\cos(\tau+\vec{k}\cdot \vec{x}_{2}+\vec{k}\cdot \vec{x}_{3})-2\mu_{21}\cos(\tau+\vec{k}\cdot \vec{x}_{1}+\vec{k}\cdot \vec{x}_{2}),
\end{eqnarray*}
\begin{eqnarray*}
    D_{4} =&& (2\mu_{13}+\mu_{23}-\mu_{21})\sin(2\tau)-4\mu_{13}\sin(\tau+\vec{k}\cdot \vec{x}_{1}+\vec{k}\cdot \vec{x}_{3})\nonumber\\
        &&{}-2\mu_{23}\sin(\tau+\vec{k}\cdot \vec{x}_{2}+\vec{k}\cdot \vec{x}_{3})+2\mu_{21}\sin(\tau+\vec{k}\cdot \vec{x}_{1}+\vec{k}\cdot \vec{x}_{2}),\\
    D_{5} =&& 2(\mu_{12}-\mu_{13}+\mu_{23})\left[ 1+\cos(2\tau) \right]-4\mu_{12}\cos(\tau+\vec{k}\cdot \vec{x}_{1}+\vec{k}\cdot \vec{x}_{2})\nonumber\\
        &&+4\mu_{13}\cos(\tau+\vec{k}\cdot \vec{x}_{1}+\vec{k}\cdot \vec{x}_{3})-4\mu_{23}\cos(\tau+\vec{k}\cdot \vec{x}_{2}+\vec{k}\cdot \vec{x}_{3}),\\
    D_{6} =&& -2(\mu_{12}-\mu_{13}+\mu_{23})\sin(2\tau)+4\mu_{12}\sin(\tau+\vec{k}\cdot \vec{x}_{1}+\vec{k}\cdot \vec{x}_{2})\nonumber\\
        &&{}-4\mu_{13}\sin(\tau+\vec{k}\cdot \vec{x}_{1}+\vec{k}\cdot \vec{x}_{3})+4\mu_{23}\sin(\tau+\vec{k}\cdot \vec{x}_{2}+\vec{k}\cdot \vec{x}_{3}).
\end{eqnarray*}

\section{Power spectral density of TDI combinations} \label{sec:appendixB}
The noise power spectral density of different TDI channels are calculated and summarized below.
Michelson X, Y, Z:
\begin{equation}
    N_{X} = 16\sin^{2}(\tau) \left\{ \left[ 3+\cos(2\tau) \right]S_{acc} + S_{oms} \right\}.
\end{equation}
Fully symmetric Sagnac $\zeta$:
\begin{equation}
    N_{\zeta} = 24\sin^{2}\left(\frac{\tau}{2} \right) \; S_{acc} + 6S_{oms}.
\end{equation}
Sagnac $\alpha$, $\beta$, $\gamma$:
\begin{equation}
    N_{\alpha} =N_{\beta}=N_{\gamma}= \left[ 16\sin^{2}\left(\frac{\tau}{2}\right) + 8\sin^{2} \left(\frac{3\tau}{2}\right) \right]  S_{acc}+ 6S_{oms}.
\end{equation}
Beacon P, Q, R:
\begin{equation}
    N_{P} =N_{Q} =N_{R} = 16\left[ 2\sin^{4}\left(\frac{\tau}{2}\right)+\sin^{2}(\tau)\right]S_{acc} + 8\left[ \sin^{2}\left(\frac{\tau}{2}\right)+\sin^{2}(\tau)\right]S_{oms}.
\end{equation}
Monitor E, F, G:
\begin{equation}
    N_{E} =N_{F} =N_{G} = 16\left[ 2\sin^{4}\left(\frac{\tau}{2}\right)+\sin^{2}(\tau)\right]S_{acc} + 8\left[ \sin^{2}\left(\frac{\tau}{2}\right)+\sin^{2}(\tau)\right]S_{oms}.
\end{equation}
Relay U, V, W:
\begin{eqnarray}
N_{U} = N_{V}=N_{W}&=&8\left[ 2\sin^{2}\left(\frac{\tau}{2}\right)+\sin^{2}(\tau)+2\sin^{2}\left(\frac{3\tau}{2}\right)\right]S_{acc} \nonumber \\ 
     &&{} + 4\left[ \sin^{2}\left(\frac{\tau}{2}\right)+2\sin^{2}(\tau)+\sin^{2}\left(\frac{3\tau}{2}\right)\right]S_{oms} .  
\end{eqnarray}
Optimal A, E, T:
\begin{eqnarray}
    N_{A} &=& N_{E} = 8\sin^{2}(\tau)\; \left\{ 2\left[3+2\cos\tau+\cos(2\tau)\right]\;S_{acc} + \left(2+\cos\tau\right)\;S_{oms} \right\}, \\
    N_{T} &=& 32\sin^{2}\left(\frac{\tau}{2}\right)\sin^{2}(\tau) \; \left[4\sin^{2}\left(\frac{\tau}{2}\right)\;S_{acc} + S_{oms} \right].
\end{eqnarray}

\bibliography{main}
\end{document}